\documentclass[10pt]{article}
\usepackage{amsmath}
\usepackage{latexsym}
\usepackage{amssymb}
\usepackage{amsfonts}
\usepackage{amsthm}
\title{AIRY, BELTRAMI, MAXWELL,  MORERA,\\ EINSTEIN AND LANCZOS POTENTIALS REVISITED}
\author{J.-F. Pommaret \\ CERMICS, Ecole des Ponts ParisTech,\\ 6/8 Av. Blaise Pascal, 77455 Marne-la-Vall\'ee Cedex 02, France \\
E-mail: jean-francois.pommaret@wanadoo.fr, pommaret@cermics.enpc.fr \\
URL: http://cermics.enpc.fr/$\sim$pommaret/home.html }
\date{  }
\begin{document}
\maketitle

\noindent
{\bf ABSTRACT}

The main purpose of this paper is to revisit the well known {\it potentials}, called {\it stress functions}, needed in order to study the parametrizations of the stress equations, respectively  provided by G.B. Airy (1863) for $2$-dimensional elasticity, then by E. Beltrami (1892), J.C. Maxwell (1870) and G. Morera (1892) for $3$-dimensional elasticity, finally by A. Einstein (1915) for $4$-dimensional elasticity, both with a variational procedure introduced by C. Lanczos (1949,1962) in order to relate potentials to Lagrange multipliers. Using the methods of {\it Algebraic Analysis}, namely mixing differential geometry with homological algebra and 
combining the {\it double duality test} involved with the {\it Spencer cohomology}, we shall be able to extend these results to an arbitrary situation with an arbitrary dimension $n$. We shall also explain why double duality is perfectly adapted to variational calculus with differential constraints as a way to eliminate the corresponding Lagrange multipliers. For example, the {\it canonical parametrization} of the stress equations is just described by the formal adjoint of the $n^2(n^2-1)/12$ components of the linearized Riemann tensor considered as a linear second order differential operator but the minimum number of potentials needed in elasticity theory is equal to $n(n-1)/2$ for any {\it minimal parametrization}. Meanwhile, we can provide all the above results without even using indices for writing down explicit formulas in the way it is done in any textbook today. The example of relativistic continuum mechanics with $n=4$ is provided in order to prove that it could be {\it strictly impossible} to obtain such results without using the above methods. We also revisit the {\it possibility} (Maxwell equations of electromagnetism) or the {\it impossibility} (Einstein equations of gravitation) to obtain canonical or minimal parametrizations for various other equations of physics. It is nevertheless important to notice that, when $n$ and the algorithms presented are known, most of the calculations can be achieved by using computers for the corresponding symbolic computations. Finally, though the paper is mathematically oriented as it aims providing new insights towards the mathematical foundations of elasticity theory and mathematical physics, it is written in a rather self-contained way.\\

\noindent
{\bf 1) INTRODUCTION}  \\

The language of {\it differential modules} has been recently introduced in control theory as a way to understand in an intrinsic way the {\it structural properties} of systems 
of ordinary differential (OD) equations (controllability, observability, identifiability, ...), but it can also be applied to systems of partial differential (PD) equations ([4],[15],[25],[26],[28],[32],[45],[46],[47],[54]). A similar comment can be done for optimal control, that is for variational calculus with differential constraints and the author thanks Prof. Lars Andersson (Einstein Institute, Potsdam) for having suggested him to study the Lanczos potential within this new framework. \\

We start providing a few explicit examples in order to convince the reader that the corresponding computations are often becoming so tricky that nobody could achieve them or even imagine any underlying general algorithm, for example in the study of the mathematical foundations of control theory, elasticity theory or general relativity. \\

\noindent
{\bf EXAMPLE 1.1}: {\it OD Control Theory} \\
With one independent variable $x$, for example the time $t$ in control theory or the curvilinear abcissa $s$ in the study of a beam, and three unknowns $(y^1, y^2, y^3)$. Setting formally $d_xy^k=y^k_x$ for $k=1,2,3$ and so on,  let us consider the system made by the two {\it first order} OD equations depending on a variable coefficient $a(x)$:  \\
\[y^3_x-a(x)y^2-y^1_x=0, \hspace{3mm} y^3-y^2_x+y^1_x=0  \]
In control theory, if $ a=cst$ is a constant parameter, one could bring the system to {\it any first order Kalman form} and check that the corresponding control system is {\it controllable} if and only if $a(a-1)\neq 0$, that is $a\neq 0$ and $a\neq 1$ (exercise), independently of the choice of $1$ {\it input} and $2$ {\it outputs} among the $3$ control variables. In addition to that, using the second OD equation in the form $y^3=y^2_x-y^1_x$ and substituting in the first, we get the only {\it second order} OD equation:  \\
\[ y^1_{xx} + y^1_x - y^2_{xx} + a(x)y^2=0  \]
a result leading to a kind of "vicious circle" because the only way to test controllability is ... to bring this second order equation back to a first order system and there are a lot of possibilities. Again, {\it in any case}, the only critical values are $a=0$ and $a=1$. Of course, one could dream about a direct approach providing the same result in an intrinsic way. Introducing the operator $d=d_x$ as the (formal) derivative with respect to $x$, we may rewrite the last equation in the form:  \\
\[      d(d + 1)y^1=(d^2 - a)y^2   \]
Replacing the operators $d(d+1)$ and $d^2-a$ by the polynomials $\chi (\chi +1)$ and ${\chi}^2-a$, the two polynomials have a common root $\chi =0 \Rightarrow a=0$ or  $\chi =-1\Rightarrow a=1$ and we find back the desired critical values but such a result is not intrinsic at all. However, we notice that, for example $a=0 \Rightarrow d((d+1)y^1-dy^2)=0$. Introducing $z'=y^1_x+y^1-y^2_x$, we get $z'_x=0$ while $a=1 \Rightarrow (d+1)(dy^1-(d-1)y^2$ that is, setting $z"=y^1_x-y^2$, we get now $z"_x+z"=0$. Calling  "{\it torsion element} " any scalar quantity made from the unknowns and their derivatives but satisfying at least one OD equation, we discover that such quantities do exist ... if and only if $a=0$ or $a=1$ (exercise). Of course, the existence of any torsion element breaks at once the controllability of the system but the converse is not evident at all, a result leading nevertheless to the feeling that {\it a control system is controllable if and only if no torsion element can be found} and such an idea can be extended "{\it mutatis mutandis} " to any system of PD equations ([32]). However, this result could be useful if and only if there is a test for checking such a property of the system.  \\

Now, using a variable parameter $a(x)$, {\it not a word of the preceding approach is left} but the concept of a torsion element still exists. Let us prove that the condition $a(a-1)\neq 0$ becomes ${\partial}_xa + a^2 - a\neq 0$ and that the computations needed are quite far from the previous ones. We ask the reader familiar with classical control theory to make his mind a few minutes (or hours !) to agree with us before going ahead by recovering himself such a differential condition. For this purpose, let us introduce the {\it formal adjoint operators} $ad(d^2+d)=d^2-d$ and $ad(d^2-a)=d^2-a$ in the inhomogeneous system:  \\
\[  \left \{  \begin{array}{rll}
(d^2-a)y & \equiv y_{xx}-ay & =u  \\
(d^2-d)y & \equiv y_{xx}-y_x & =v
\end{array}  \right.  \]
Substracting, we get $(d-a)y\equiv y_x-ay=u-v$ and thus $y_{xx}-ay_x-{\partial}_xay=u_x-v_x$, a result leading to:  \\
\[  ({\partial}_xa+a^2-a)y=v_x-u_x+av+(1-a)u  \]
and to two possibilities:  \\

\noindent
$\bullet \hspace{3mm} \fbox{ $ {\partial}_xa+a^2-a\neq 0$}$\\

Finding $y$ and substituting it into $y_{xx}-ay=u$, we get two {\it third order} operators $P$ and $Q$ such that, among the CC of the previous system, we have $Pu-Qv=0$. Using the adjoint, we obtain therefore two {\it identities} in the operator sense:  \\
\[ P(d^2-a)\equiv Q(d^2-d)  \Rightarrow (d^2-a)ad(P)\equiv (d^2+d)ad(Q)  \]
Writing $y^1=(d^2+d)^{-1}(d^2-a)y^2$ in a symbolic way and setting $y^2=ad(P)\xi$, we obtain finally:  \\
\[  y^1=(d^2+d)^{-1}(d^2-a)ad(P)\xi=(d^2+d)^{-1}(d^2+d)ad(Q)\xi=ad(Q)\xi  \]
and discover the true reason for introducing the adjoint operator in order to invert the position of the factors. It is easy to check that we have obtained a {\it third order} parametrization of the control system. Of course, the situation $a=cst$ is much simpler as we can choose $P=d^2-d, Q=d^2-a$ and get the {\it second order} parametrization $y^1=(d^2-a)\xi,\hspace{3mm}y^2=(d^2+d)\xi$ which is easily seen to be injective, in a coherent way with controllability (exercise).\\

However, the aim of this example is to point out  the fact that {\it things are in fact quite more tricky indeed} because, substituting $y$ into into $y_{xx}-ay=u$, after a painful computation largely simplified by assuming that ${\partial}_xa+a^2-a=1$, we obtain the {\it third order} OD equation:  \\
\[   v_{xxx}+av_{xx}+(2{\partial}_xa-a)v_x+({\partial}_{xx}a-a^2)v=u_{xxx}+(a-1)u_{xx}+(2{\partial}_xa-a)u_x+({\partial}_{xx}a+{\partial}_xa)u\]
Under the same assumption, using now the OD equation $y_x-ay=v-u$ already obtained by substraction, we only get the "simpler " {\it second order} OD equation $P'(d^2-a)=Q'(d^2-d)$ in the form:  \\
\[   v_{xx}+(2{\partial}_xa-a)v=u_{xx}-u_x+2{\partial}_xau  \]
which is leading, exactly as before, to the new {\it second order} parametrization:  \\
\[   y^1=ad(Q'){\xi}'=(d^2+2{\partial}_xa-a){\xi}', \hspace{5mm}  y^2=ad(P'){\xi}'=(d^2+d+2{\partial}_xa){\xi}'  \]
which is injective as we obtain easily $y^1_x-y^2_x+(1-a)y^1+ay^2=({\partial}_xa+a^2-a){\xi}'={\xi}'$ and we let the reader check the identity:  \\
\[   (d^2+d)(d^2+2{\partial}_xa-a)\equiv (d^2-a)(d^2+d+2{\partial}_xa)   \]
The reason is that the third order CC already considered are in fact generated by the above only second order CC as it can be seen by applying the operator $d+a$ to this CC. We have indeed:   \\
\[   P=(d+a)P', Q=(d+a)Q'  \Rightarrow ad(P)=ad(P')(d-a), ad(Q)=ad(Q')(d-a)  \]
and it just remains to set ${\xi}'=(d-a)\xi$ in order to get coherent parametrizations. As a byproduct, {\it the third order parametrization is not injective} as it should only lead to $ {\xi}'=0$ and thus to ${\xi}_x-a\xi=0$. Needless to say that, if the situation is already tricky for OD equations, it should become worst for PD equations as we shall see in the next examples.  \\

\noindent
$\bullet \hspace{3mm} \fbox{ ${\partial}_xa+a^2-a=0$}$\\

Multiplying the control system by a test function $\lambda$ and integrating by parts, the kernel of the operator thus obtained is defined by the OD equations:  \\
\[ {\lambda}_{xx}-{\lambda}_x=0, \hspace{2mm}  {\lambda}_{xx} - a \lambda=0 \hspace{2mm} \Rightarrow {\lambda}_x - a  \lambda =0\hspace{2mm} \Rightarrow ({\partial}_xa+a^2-a)\lambda =0 \]
The formal adjoint of the operator defining the control system is thus no longer injective but we can repeat the previous computations in order to find the only generating CC $ v_x+av=u_x+(a-1)u $ that should lead to the injective parametrization $y^1={\xi}_x - a\xi, \hspace{2mm} y^2={\xi}_x+(1- a)\xi$ because it gives by substraction $y^2-y^1=\xi$. However, {\it such a procedure is no longer working} because we get by substitution the first order OD equation:  \\
\[  z \equiv y^2_x-ay^2-ay^2-y^1_x+(a-1)y^1=0  \]
We notice that $z$ is  a torsion element as it satisfies $z_x+az=0$ under the assumptions made on $a$.

We finally point out that such a situation cannot be found when a parametrization is existing because we may substitute it into the torsion element and get a contradiction as this procedure should provide at least one OD or PD equation among the {\it arbitrary functions} or {\it potentials} used for the parametrization.  \\

\noindent
{\bf EXAMPLE 1.2}: {\it OD Optimal Control Theory}  \\
OD optimal control is the study of OD variational calculus with OD constraints described by OD control systems. However, while studying optimal control, the author of this paper has been surprised to discover that, {\it in all cases}, the OD constraints were defined by means of controllable control systems. It is only at the end of this paper that  the importance of such an assumption will be explained. For the moment, we shall provide an example allowing to exhibit all the difficulties involved. For this, let $y^1=f^1(x), y^2=f^2(x)$ be a solution of the following single input/single output (SISO) OD control system where $a$ is a constant parameter:  \\
\[ y^1_x+y^1-y^2_x-ay^2=0  \]
Proceeding as before, the two polynomials replacing the respective operators are $\chi +1$, $\chi +a$ and can only have the common root $a=1$. Accordingly, the system is controllable if and only if $a\neq 1$ for any choice of input and output. Now, let us introduce the so-called " {\it cost function} " and let us look at the extremum of the integral $\int \frac{1}{2}((y^1)^2-(y^2)^2)dx$ under the previous OD constraint.It is well known that the proper way to study such a problem is to introduce a {\it Lagrange multiplier} $\lambda$ and to vary the new integral:  \\
\[  \int [\frac{1}{2}((y^1)^2-(y^2)^2) + \lambda (y^1_x+y^1 -y^2_x - ay^2)]dx \]
The corresponding {\it Euler-Lagrange} (EL) equations are:  \\
\[  \left \{  \begin{array}{rccl}
y^1 & \rightarrow & -{\lambda}_x+\lambda+y^1 & =0 \\
y^2 & \rightarrow &\hspace{2mm} {\lambda}_x -a \lambda-y^2 & = 0
\end{array} \right.\]
to which we must add the OD constraint when varying $\lambda$. Summing the two EL equations, we get $(a-1)\lambda=y^1-y^2$ and {\it two possibilities}:  \\
1) $a=1 \Rightarrow y^1-y^2=0$ compatible with the constaint.  \\
2) $a\neq 1 \Rightarrow \lambda=(y^1-y^2)/(a-1)$.  \\
Substituting, we get:  \\
\[ \left \{  \begin{array}{rl}
y^1_x-y^2_x-ay^1+y^2 & =0 \\
y^1_x-y^2_x+y^1-ay^2 & =0
\end{array}  \right. \]
{\it This system may not be formally integrable}. Indeed, by substraction, we get $(a+1)(y^1-y^2)=0$ and must consider the following {\it two possibilities}:  \\
\[ \left \{ \begin{array}{rccl}
\bullet & a=-1 &\Rightarrow & y^1_x+y^1-y^2_x+y^2=0  \\
\bullet & a\neq -1 & \Rightarrow & y^1=y^2=0  
\end{array}  \right.  \]
Summarising the results so far obtained, we discover that {\it the Lagrange multiplier is known if and only if the system is controllable}. Also, if $a=-1$, we may exhibit the parametrization ${\xi}_x-\xi=y^1,{\xi}_x+\xi=y^2$ and the cost function becomes $2\xi{\xi}_x=d_x({\xi}^2)$. Equivalently, {\it when the system is controllable it can be parametrized and the variational problem with constraint becomes a variational problem without constraint} which, {\it sometimes}, does not provide EL equations. We finally understand that extending such a situation to PD variational calculus with PD constraints needs new techniques.  \\

\noindent
{\bf EXAMPLE 1.3}: {\it Classical Elasticity Theory} \\
In classical elasticity, the {\it stress} tensor density $\sigma=({\sigma}^{ij}={\sigma}^{ji})$ existing inside an elastic body is a symmetric $2$-tensor density intoduced by A. Cauchy in 1822. The corresponding Cauchy {\it stress equations} can be written as ${\partial}_r{\sigma}^{ir}=f^i$ where the right member describes the local density of forces applied to the body, for example gravitation. With zero second member, we study the possibility to "{\it parametrize} " the system of PD equations 
${\partial}_r{\sigma}^{ir}=0$, namely to express its general solution by means of a certain number of arbitrary functions or {\it potentials}, called {\it stress functions}. Of course, the problem is to know about the number of such functions and the order of the parametrizing operator. In what follows, the space has $n$ local coordinates $x=(x^i)=(x^1, ... , x^n)$. For $n=1,2,3$ one may introduce the Euclidean metric $\omega=({\omega}_{ij}={\omega}_{ji})$ while, for $n=4$, one may consider the Minkowski metric. A few definitions used thereafter will be provided later on.\\

\noindent
$\bullet \hspace{3mm}n=1$: There is no possible parametrization of ${\partial}_x\sigma=0$.  \\

\noindent
$\bullet \hspace{3mm}n=2$: The stress equations become ${\partial}_1{\sigma}^{11}+{\partial}_2{\sigma}^{12}=0, {\partial}_1{\sigma}^{21}+{\partial}_2{\sigma}^{22}=0$. Their second order parametrization ${\sigma}^{11}={\partial}_{22}\phi, {\sigma}^{12}={\sigma}^{21}=-{\partial}_{12}\phi, {\sigma}^{22}={\partial}_{11}\phi$ has been provided by George Biddell Airy (1801-1892) in 1863 ([1]). It can be recovered by simply rewriting the stress equations in the following manner: \\
\[ {\partial}_1{\sigma}^{11}={\partial}_2( - {\sigma}^{12})\Rightarrow \exists \varphi, {\sigma}^{11}={\partial}_2\varphi, {\sigma}^{12}= - {\partial}_1\varphi,   {\partial}_2{\sigma}^{22}={\partial}_1( - {\sigma}^{21})\Rightarrow \exists \psi, {\sigma}^{22}={\partial}_1\psi, {\sigma}^{21}= - {\partial}_2\psi \] 
\[   {\sigma}^{12}={\sigma}^{21} \Rightarrow {\partial}_1 \varphi={\partial}_2\psi \Rightarrow \exists \phi, \varphi={\partial}_2\phi, \psi={\partial}_1\phi  \]
We get the second order system:  \\
\[ \left\{  \begin{array}{rll}
{\sigma}^{11} & \equiv {\partial}_{22}\phi =0 \\
-{\sigma}^{12} & \equiv {\partial}_{12}\phi =0 \\
{\sigma}^{22} & \equiv {\partial}_{11}\phi=0
\end{array}
\right. \fbox{ $ \begin{array}{ll}
1 & 2   \\
1 & \bullet \\  
1 & \bullet  
\end{array} $ } \]
which is involutive with one equation of class $2$, $2$ equations of class $1$ and it is easy to check that the $2$ corresponding first order CC are just the stress equations.\\
As we have a system with constant coefficients, we may use localization in order to transform the $2$ PD equations into the $2$ {\it linear} equations ${\chi}_1{\sigma}^{11}+{\chi}_2{\sigma}^{12}=0,{\chi}_1{\sigma}^{21}+{\chi}_2{\sigma}^{22}=0$ and get\\
\[  {\sigma}^{11}=-\frac{{\chi}_2}{{\chi}_1}{\sigma}^{12}=-\frac{({\chi}_2)^2}{{\chi}_1{\chi}_2}{\sigma}^{12},\hspace{4mm}
 {\sigma}^{22}=-\frac{{\chi}_1}{{\chi}_2}{\sigma}^{12}=-\frac{({\chi}_1)^2}{{\chi}_1{\chi}_2}{\sigma}^{12} \]
Setting ${\sigma}^{12}= - {\chi}_1{\chi}_2 \phi$, we finally get ${\sigma}^{11}=({\chi}_2)^2\phi, {\sigma}^{22}=({\chi}_1)^2 \phi$ and obtain the previous parametrization by delocalizing, that is replacing now ${\chi}_i$ by ${\partial}_i$. \\

\noindent
$\bullet \hspace{3mm} n=3$: Things become quite more delicate when we try to parametrize the $3$ PD equations: \\
\[ {\partial}_1{\sigma}^{11}+{\partial}_2{\sigma}^{12}+{\partial}_3{\sigma}^{13}=0,\hspace{3mm} {\partial}_1{\sigma}^{21}+{\partial}_2{\sigma}^{22}+{\partial}_3{\sigma}^{23}=0, \hspace{3mm} {\partial}_1{\sigma}^{31}+{\partial}_2{\sigma}^{32}+{\partial}_3{\sigma}^{33}=0 \]
Of course, localization could be used similarly by dealing with the $3$ {\it linear} equations:  \\
\[ {\chi}_1{\sigma}^{11}+{\chi}_2{\sigma}^{12}+{\chi}_3{\sigma}^{13}=0, {\chi}_1{\sigma}^{21}+{\chi}_2{\sigma}^{22}+{\chi}_3{\sigma}^{23}=0, 
{\chi}_1{\sigma}^{31}+{\chi}_2{\sigma}^{32}+{\chi}_3{\sigma}^{33}=0 \]
having rank $3$ for $6$ unknowns but, even if we succeed bringing all the fractions to the same denominator as before after easy but painful calculus, there is an additional difficulty which is well hidden. Indeed, coming back to the previous Example when $a=cst$, say $a=1$, we should get $(d^2+d)y^1=(d^2-1)y^2 \Rightarrow \chi(\chi +1)y^1=(\chi +1)(\chi -1)y^2\Rightarrow \chi y^1=(\chi -1)y^2 \Rightarrow y^1=\frac{\chi -1}{\chi}y^2$. Hence, setting $y^2=\chi y, y^1=(\chi -1)y$, we only get a parametrization of the {\it first order} OD equation $z\equiv {\dot{y}}^1-{\dot{y}}^2+y^2=0$ leading to $\dot{z}+z=0$. Accordingly, localization does indeed provide a parametrization, ... if we already know there exists a possibility to parametrize the given system or if we are able to check that we have obtained such a parametrization by using involution, a way to supersede the use of Janet or Gr\"{o}bner bases as was proved for the case $n=2$ ([34]). Also, if we proceed along such a way, we should surely loose any geometric argument that could stay behind such a procedure, ... if there is one !.\\

A direct computational approach has been provided by Eugenio Beltrami (1835-1900) in 1892 ([3]), James Clerk Maxwell (1831-1879) in 1870 ([23]) and Giacinto Morera (1856-1909) in 1892 ([24]) by introducing the $6$ stress functions ${\phi}_{ij}={\phi}_{ji}$ through the parametrization obtained by considering:\\
\[{\sigma}^{11} = {\partial}_{33}{\phi}_{22}+{\partial}_{22}{\phi}_{33}-2{\partial}_{23}{\phi}_{23} \]
\[{\sigma}^{12}={\sigma}^{21} = {\partial}_{13}{\phi}_{23}+{\partial}_{23}{\phi}_{13}-{\partial}_{33}{\phi}_{12}-{\partial}_{12}{\phi}_{33}  \]
and the additional $4$ relations obtained by using a cyclic permutation of $(1,2,3)$. The corresponding system:\\
\[   \left\{  \begin{array}{rll}
{\sigma}^{11} \equiv& {\partial}_{33}{\phi}_{22}+{\partial}_{22}{\phi}_{33}-2{\partial}_{23}{\phi}_{23}=0  \\
-{\sigma}^{12}\equiv & {\partial}_{33}{\phi}_{12}+{\partial}_{12}{\phi}_{33}-{\partial}_{13}{\phi}_{23}-{\partial}_{23}{\phi}_{13}=0  \\
 {\sigma}^{22}\equiv & {\partial}_{33}{\phi}_{11}+{\partial}_{11}{\phi}_{33}-2{\partial}_{13}{\phi}_{13}=0  \\
{\sigma}^{13}\equiv & {\partial}_{23}{\phi}_{12}+{\partial}_{12}{\phi}_{23}-{\partial}_{22}{\phi}_{13}-{\partial}_{13}{\phi}_{22} =0 \\
-{\sigma}^{23}\equiv & {\partial}_{23}{\phi}_{11}+{\partial}_{11}{\phi}_{23}-{\partial}_{12}{\phi}_{13}-{\partial}_{13}{\phi}_{12} =0 \\
{\sigma}^{33}\equiv & {\partial}_{22}{\phi}_{11}+{\partial}_{11}{\phi}_{22}-2{\partial}_{12}{\phi}_{12}=0
\end{array}
\right. \fbox{ $ \begin{array}{lll}
1 & 2 & 3   \\
1 & 2 & 3  \\
1 & 2 & 3  \\
1 & 2 &  \bullet  \\
1 & 2 & \bullet  \\
1 & 2 & \bullet
\end{array} $ } \]
is involutive with $3$ equations of class $3$, $3$ equations of class $2$ and no equation of class $1$. The $3$ CC are describing the stress equations which admit therefore a parametrization ... justifying the localization approach "{\it a posteriori} " but without any geometric framework.\\

{\it Surprisingly}, the Maxwell parametrization is obtained by keeping ${\phi}_{11}=A, {\phi}_{22}=B, {\phi}_{33}=C$ while setting ${\phi}_{12}={\phi}_{23}={\phi}_{31}=0$ in order to obtain the system:\\
\[   \left\{  \begin{array}{rl}
{\sigma}^{11} \equiv& {\partial}_{33}B+{\partial}_{22}C=0  \\
{\sigma}^{22}\equiv & {\partial}_{33}A+{\partial}_{11}C =0 \\
- {\sigma}^{23}\equiv & {\partial}_{23}A=0  \\
{\sigma}^{33}\equiv & {\partial}_{22}A+{\partial}_{11}B=0  \\
- {\sigma}^{13}\equiv &{\partial}_{13}B=0 \\
- {\sigma}^{12}\equiv & {\partial}_{12}C=0
\end{array}
\right. \fbox{ $ \begin{array}{lll}
1 & 2 & 3   \\
1 & 2 & 3  \\
1 & 2 & \bullet  \\
1 & 2 &  \bullet  \\
1 & \bullet & \bullet  \\
1 & \bullet & \bullet
\end{array} $ } \]
However, {\it this system may not be involutive} and no CC can be found "{\it a priori} " because the coordinate system is surely not $\delta$-regular. Indeed, effecting the linear change of coordinates $x^1 \rightarrow x^1+x^3, x^2\rightarrow x^2+x^3, x^3\rightarrow x^3$, we obtain the involutive system:  \\
\[   \left\{  \begin{array}{l}
{\partial}_{33}C+{\partial}_{13}C+{\partial}_{23}C+{\partial}_{12}C=0  \\
{\partial}_{33}B+{\partial}_{13}B=0  \\
{\partial}_{33}A+{\partial}_{23}A=0  \\
{\partial}_{23}C-{\partial}_{13}B-{\partial}_{13}C-{\partial}_{12}C+{\partial}_{22}C=0  \\
{\partial}_{23}A-{\partial}_{22}C+{\partial}_{13}B+2{\partial}_{12}C-{\partial}_{11}C=0  \\
{\partial}_{22}A+{\partial}_{22}C-2{\partial}_{12}C+{\partial}_{11}B+{\partial}_{11}C=0
\end{array}
\right. \fbox{ $ \begin{array}{lll}
1 & 2 & 3   \\
1 & 2 & 3  \\
1 & 2 &  3  \\
1 & 2 &  \bullet  \\
1 &  2 & \bullet  \\
1 &  2 & \bullet
\end{array} $ } \]
and it is easy to check that the $3$ CC obtained just amount to the desired $3$ stress equations when coming back to the original system of coordinates. Again, if there is a geometrical background, this change of local coordinates is hidding it totally. Moreover, we notice that the stress functions kept in the procedure are just the ones on which ${\partial}_{33}$ is acting. The reason for such an apparently technical choice is related to very general deep arguments in the theory of differential modules that will only be explained at the end of the paper.\\

Finally, the Morera parametrization is obtained by keeping now ${\phi}_{23}=L, {\phi}_{13}=M, {\phi}_{12}=N$ while setting ${\phi}_{11}={\phi}_{22}={\phi}_{33}=0$ in order  to obtain the system:  \\
\[   \left\{  \begin{array}{rl}
{\sigma}^{11}\equiv & -2 {\partial}_{23}L=0  \\
{\sigma}^{22}\equiv & -2{\partial}_{13}M=0  \\
{\sigma}^{33}\equiv & -2{\partial}_{12}N=0  \\
{\sigma}^{12}\equiv &{\partial}_{13}L+{\partial}_{23}M-{\partial}_{33}N=0 \\
{\sigma}^{23}\equiv & {\partial}_{12}M+{\partial}_{13}N-{\partial}_{11}L=0  \\
{\sigma}^{13}\equiv & {\partial}_{23}N+{\partial}_{12}L-{\partial}_{22}M=0
\end{array}  
\right.  \]
which is involutive because convenient $\delta$-regular coordinates can be similarly exhibited and provide again the $3$ desired stress equations (exercise)
(Compare to [53]). \\

\noindent
$\bullet \hspace{3mm}n\geq 4$: As already explained, localization {\it cannot} be applied directly as we don't know if a parametrization may exist and in any case no analogy with the previous situations $n=1,2,3$ could be used. Moreover, no known differential geometric background could be used at first sight in order to provide a hint towards the solution. Now, if $\omega$ is the Minkowski metric and $\phi=GM/r$ is the gravitational potential, then $\phi/c^2\ll 1$ and a perturbation $\Omega\in S_2T^*$  of $\omega$ may satisfy in vacuum the $10$ second order {\it Einstein equations} for the $10$ $ \Omega$:  \\
\[   E_{ij}\equiv {\omega}^{rs}(d_{ij}{\Omega}_{rs}+d_{rs}{\Omega}_{ij}-d_{ri}{\Omega}_{sj}-d_{sj}{\Omega}_{ri})-{\omega}_{ij}({\omega}^{rs}{\omega}^{uv}d_{rs}{\Omega}_{uv}
-{\omega}^{ru}{\omega}^{sv}d_{rs}{\Omega}_{uv})=0  \]
by introducing the corresponding second order {\it Einstein operator} $S_2T^* \stackrel{Einstein}{\longrightarrow} S_2T^*:\Omega \rightarrow E$ when $n=4$ ([11]). Though it is well known that the corresponding second order {\it Einstein operator} is parametrizing the stress equations, the challenge of parametrizing Einstein equations has been proposed in 1970 by J. Wheeler for 1000 \$ and solved {\it negatively} in 1995 by the author who only received 1 \$. We shall see that, {\it exactly as before and though it is quite striking}, the key ingredient will be to use the linearized Riemann tensor considered as a second order operator ([29],[32]). As an {\it even more striking fact}, we shall discover that the condition $n\geq 4$ has only to do with Spencer cohomology for the symbol of the {\it conformal Killing equations}.\\

\noindent
{\bf EXAMPLE 1.4}: {\it PD Control Theory} \\
The aim of this last example is to prove that the possibility to exhibit two different parametrizations of the stress equations which has been presented in the previous example has surely nothing to do with the proper mathematical background of elasticity theory !.  \\
For this, let us consider the (trivially involutive) inhomogeneous PD equations with two independent variables $(x^1,x^2)$, two unknown functions $({\eta}^1, {\eta}^2)$ and a second member $\zeta$: \\
\[   {\partial}_2{\eta}^1 - {\partial}_1{\eta}^2 + x^2 {\eta}^2=\zeta   \]
Multiplying on the left by a test function $\lambda$ and integrating by parts, the corresponding inhomogeneous system of PD equations is:  \\
\[  \left\{ \begin{array}{rcll}
{\eta}^1 & \rightarrow  & - {\partial}_2 \lambda  & ={\mu}^1  \\
{\eta}^2 & \rightarrow & \hspace{3mm}{\partial}_1 \lambda + x^2 \lambda  & ={\mu}^2
\end{array} \right.  \]
Using crossed derivatives, we get $\lambda={\partial}_2{\mu}^2+{\partial}_1{\mu}^1+x^2{\mu}^1$ and substituting, we get the two CC:  \\
\[  \left  \{  \begin{array}{lcl}
- {\partial}_{22}{\mu}^2 - {\partial}_{12}{\mu}^1-x^2{\partial}_2{\mu}^1 - 2{\mu}^1  & = & {\nu}^1\\
\hspace{3mm} {\partial}_{12}{\mu}^2 + {\partial}_{11}{\mu}^1+2x^2{\partial}_1{\mu}^1+x^2{\partial}_2{\mu}^2+(x^2)^2{\mu}^1-{\mu}^2 & = & {\nu}^2  
\end{array}  \right.  \]
The corresponding generating CC for the second member $({\nu}^1,{\nu}^2)$ is:  \\
\[  {\partial}_2{\nu}^2 + {\partial}_1{\nu}^1 + x^2 {\nu}^1=0  \]
Therefore ${\nu}^2$ is differentially dependent on ${\nu}^1$ but ${\nu}^1$ is also differentially dependent on ${\nu}^2$.\\
Multiplying the first equation by the test function ${\xi}^1$, the second equation by the test function ${\xi}^2$, adding and integrating by parts, we get the canonical parametrization $({\xi}^1,{\xi}^2)\rightarrow ({\eta}^1,{\eta}^2)$:  \\
\[   \left \{\begin{array}{rcll}
{\mu}^2 & \rightarrow & -{\partial}_{22}{\xi}^1+{\partial}_{12}{\xi}^2-x^2{\partial}_2{\xi}^2-2{\xi}^2  & = {\eta}^2  \\
{\mu}^1&  \rightarrow & -{\partial}_{12}{\xi}^1+x^2{\partial}_2{\xi}^1-{\xi}^1+{\partial}_{11}{\xi}^2  -2x^2{\partial}_1{\xi}^2+(x^2)^2{\xi}^2 & = {\eta}^1        
\end{array}  \right. \fbox{ $ \begin{array}{ll}
1 & 2   \\
1 & \bullet 
\end{array} $ } \]
of the initial system with zero second member. The system (up to sign) is involutive and the kernel of this parametrization has differential rank equal to $1$.\\
Keeping ${\xi}^1=\xi$ while setting ${\xi}^2=0$, we get the first minimal parametrization $\xi \rightarrow ({\eta}^1,{\eta}^2)$:  \\
\[  \left  \{  \begin{array}{ll}
-{\partial}_{22}\xi & = {\eta}^2  \\
-{\partial}_{12}\xi+x^2{\partial}_2\xi-\xi & ={\eta}^1
\end{array}  \right.  \fbox{ $ \begin{array}{ll}
1 & 2   \\
1 & \bullet 
\end{array} $ }  \]
The system is again involutive (up to sign) and the parametrization is minimal because the kernel of this parametrization has differential rank equal to $0$.\\
Setting now ${\xi}^1=0$ while keeping ${\xi}^2={\xi}'$, we get the second minimal parametrization ${\xi}' \rightarrow ({\eta}^1,{\eta}^2)$:  \\
\[  \left  \{\begin{array}{ll}
{\partial}_{11}{\xi}'-2x^2{\partial}_1{\xi}'+(x^2)^2{\xi}' & = {\eta}^1  \\
{\partial}_{12}{\xi}'  - x^2{\partial}_2{\xi}' - 2{\xi}' & = {\eta}^2 
\end{array}   \right.  \]
with a similar comment.  \\

\noindent
{\bf EXAMPLE 1.5}: {\it PD Optimal Control Theory} \\
   Let us revisit briefly the foundation of n-dimensional elasticity theory as it can be found today in any textbook, restricting our study to $n=2$ for simplicity. If $x=(x^1,x^2)$ is a point in the plane and $\xi=({\xi}^1(x),{\xi}^2(x))$ is the displacement vector, lowering the indices by means of the Euclidean metric, we may introduce the "small" deformation tensor $\epsilon = ({\epsilon}_{ij}={\epsilon}_{ji}=(1/2)({\partial}_i{\xi}_j+{\partial}_j{\xi}_i))$ with $n(n+1)/2=3$ (independent) {\it components} $({\epsilon}_{11}, {\epsilon}_{12}={\epsilon}_{21},{\epsilon}_{22})$. If we study a part of a deformed body, for example a thin elastic plane sheet, by means of a variational principle, we may introduce the local density of free energy $\varphi (\epsilon)=\varphi({\epsilon}_{ij}{\mid} i\leq j)=\varphi ({\epsilon}_{11},{\epsilon}_{12},{\epsilon}_{22})$ and vary the total free energy $\Phi={\int}\varphi (\epsilon)dx$ with $dx=dx^1\wedge dx^2$ by introducing ${\sigma}^{ij}=\partial \varphi/\partial {\epsilon}_{ij}$ for $i\leq j$ in order to obtain $\delta \Phi={\int}({\sigma}^{11}\delta {\epsilon}_{11}+{\sigma}^{12}\delta {\epsilon}_{12}+{\sigma}^{22}\delta{\epsilon}_{22})dx$. Accordingly, {\it the "decision" to define the stress tensor} $\sigma$ {\it by a symmetric matrix with} ${\sigma}^{12}={\sigma}^{21}$ {\it is purely artificial within such a variational principle}. Indeed, the usual Cauchy device (1828) assumes that each element of a boundary surface is acted on by a surface density of force ${\vec{\sigma}}$ with a linear dependence $\vec{\sigma}=({\sigma}^{ir}(x)n_r)$ on the outward normal unit vector $\vec{n}=(n_r)$ and does not make any assumption on the stress tensor. It is only by an equilibrium of forces and couples, namely the well known {\it phenomenological static torsor equilibrium}, that one can " {\it prove} " the symmetry of $\sigma$. However, even if we assume this symmetry, {\it we now need the different summation} ${\sigma}^{ij}\delta{\epsilon}_{ij}= {\sigma}^{11}\delta{\epsilon}_{11}+2{\sigma}^{12}\delta{\epsilon}_{12}+{\sigma}^{22}\delta{\epsilon}_{22}={\sigma}^{ir}{\partial}_r\delta {\xi}_i$. An integration by parts and a change of sign produce the 
 integral ${\int}({\partial}_r{\sigma}^{ir})\delta {\xi}_idx$ leading to the stress equations ${\partial}_r{\sigma}^{ir}=0$ already considered. This classical approach to elasticity theory, based on invariant theory with respect to the group of rigid motions, cannot therefore describe equilibrium of torsors by means of a variational principle where the proper torsor concept is totally lacking. It is however widely used through the technique of " {\it finite elements} " where it can also be applied to electromagnetism (EM) with similar quadratic (piezoelectricity) or cubic (photoelasticity) lagrangian integrals. In this situation, the $4$-potential $A$ of EM is used in place of $\xi$ while the EM field $dA=F=(\vec{B}, \vec{E})$ is used in place of $\epsilon$ and the Maxwell equations $dF=0$ are used in place of the Riemann CC for 
 $\epsilon$. \\
 
However, there exists another equivalent procedure dealing with a {\it variational calculus with constraint}. Indeed, as we shall see later on, the deformation tensor is not any symmetric tensor as it must satisfy $n^2(n^2-1)/12$ compatibility conditions (CC), that is only ${\partial}_{22}{\epsilon}_{11}+{\partial}_{11}{\epsilon}_{22}-2{\partial}_{12}{\epsilon}_{12}=0$ when $n=2$. In this case, introducing the {\it Lagrange multiplier } $\lambda$, {\it we have to vary the new integral} $\int[{\varphi}(\epsilon) + \lambda ({\partial}_{22}{\epsilon}_{11}+{\partial}_{11}{\epsilon}_{22}-2{\partial}_{12}{\epsilon}_{12})]dx$ {\it for an arbitrary} $\epsilon$. Setting $\lambda= - \phi$, a double integration by parts now provides the parametrization ${\sigma}^{11}={\partial}_{22}\phi,{\sigma}^{12}={\sigma}^{21}=-{\partial}_{12}\phi, {\sigma}^{22}={\partial}_{11}\phi$ of the stress equations by means of the Airy function $\phi$ and the {\it formal adjoint} of the Riemann CC, {\it on the condition to observe that we have in fact} $2{\sigma}^{12}=-2{\partial}_{12}\phi$ as another way to understand the deep meaning of the factor "2" in the summation. The same variational calculus with constraint may thus also be used in order to " shortcut " the introduction of the EM potential. \\
Finally, using the {\it constitutive relations} of the material establishing an isomorphism $\sigma \longleftrightarrow \epsilon$, one can also introduce a local free energy $\psi (\sigma)$ in a similar variational problem having for constraint the stress equations, with the same comment as above (See [32], p 915, for more details). The well known {\it Minkowski constitutive relations} $(\vec{B},\vec{E}) \longleftrightarrow (\vec{H},\vec{D})$ can be similarly used for EM.  \\

In arbitrary dimension, the above compatibility conditions are nothing else but the linearized Riemann tensor in Riemannian geometry, a crucial mathematical tool in the theory of general relativity and a good reason for studying the work of Cornelius Lanczos (1893-1974) as it can be found in ([18],[19],[20]) or in a few modern references ([2],[7]-[10],[22],[27],[49]). The starting point of Lanczos has been to take EM as a model in order to introduce a Lagrangian that should be quadratic in the Riemann tensor $({\rho}^k_{l,ij}\Rightarrow {\rho}_{ij}={\rho}^r_{i,rj}={\rho}_{ji}\Rightarrow \rho={\omega}^{ij}{\rho}_{ij})$ while considering it independently of its expression through the second order derivatives of a metric $({\omega}_{ij})$ with inverse $({\omega}^{ij})$ or the first order derivatives of the corresponding Christoffel symbols $({\gamma}^k_{ij})$. According to the previous paragraph, the corresponding variational calculus {\it must} involve PD constraints made by the Bianchi identities and the new lagrangian to vary must therefore contain as many Lagrange multipliers as the number of Bianchi identities that can be written under the form:  
\[   {\nabla}_r{\rho}^k_{l,ij}+{\nabla}_i{\rho}^k_{l,jr}+{\nabla}_j{\rho}^k_{l,ri}=0 \Rightarrow   {\nabla}_r{\rho}^r_{l,ij}={\nabla}_i{\rho}_{lj}-{\nabla}_j{\rho}_{li}       \]
Meanwhile, Lanczos and followers have been looking for a kind of parametrization of the Bianchi identities, exactly like the Lagrange multiplier has been used as an Airy potential for the stress equations. However, we shall prove that the definition of a {\it Riemann candidate} and the answer to this question cannot be done without the knowledge of the Spencer cohomology. Moreover, we have pointed out the existence of well known couplings between elasticity and electromagnetism, namely piezoelectricity and photoelasticity, which are showing that, in the respective Lagrangians, the EM field is on equal footing with the deformation tensor and {\it not} with the Riemann tensor. This fact is showing the {\it shift by one step} that must be used in the physical interpretation of the differential sequences involved and cannot be avoided. Meanwhile, the {\it ordinary derivatives} ${\partial}_i$ can be used in place of the {\it covariant derivatives} ${\nabla}_i$ when dealing with the linearized framework as the Christoffel symbols vanish when Euclidean or Minkowskian metrics are used. \\
The next tentative of Lanczos has been to extend his approach to the Weyl tensor:  \\
\[  {\tau}^k_{l,ij} = {\rho}^k_{l,ij} - \frac{1}{(n-2)}({\delta}^k_i{\rho}_{lj} - {\delta}^k_j{\rho}_{li} +{\omega}_{lj}{\omega}^{ks}{\rho}_{si} - {\omega}_{li}{\omega}^{ks}{\rho}_{sj}) + \frac{1}{(n-1)(n-2)}({\delta}^k_i{\omega}_{lj} - {\delta}^k_j{\omega}_{li})\rho  \]
The main problem is now that the Spencer cohomology of the symbols of the conformal Killing equations, in particular the $2$-acyclicity, will be {\it absolutely 
needed} in order to study the Vessiot structure equations providing the Weyl tensor and its relation with the Riemann tensor. It will follow that the CC for the Weyl tensor are not first order contrary to the CC for the Riemann tensor made by the Bianchi identities, another reason for justifying the shift by one step already quoted. In order to provide an idea of the difficulty involved, let us define the following tensors:  \\
\[     Schouten= ({\sigma}_{ij}={\rho}_{ij} - \frac{1}{2(n-1)}{\omega}_{ij}\rho) \Rightarrow Cotton=({\sigma}_{k,ij}={\nabla}_i{\sigma}_{kj} - {\nabla}_j{\sigma}_{ki})  \]
An elementary but tedious computation allows to prove the formula:  \\
\[              {\nabla}_r{\tau}^r_{k,ij}=\frac{(n-3)}{(n-2)}{\sigma}_{k,ij}   \]
Then, of course, {\it if Einstein equations in vacuum are valid}, the Schouten and Cotton tensors vanish but the left member is by no way a differential identity for the Weyl tensor as care must be taken when mixing up mathematics with physics.  \\

Finally, comparing the various parametrizations already obtained in the previous examples, it seems that the procedures are similar, even when dealing with 
systems having variable coefficients. The purpose of the paper is to prove that, in order to obtain a general algorithm, we shall need a lot of new tools involving at the same time {\it commutative algebra, homological algebra, differential algebra} and {\it differential geometry} that will be recalled in the next sections. Moreover, we want to point out the fact that {\it the use of differential modules is necessary} as it is the only possibility to avoid any functional background like the concept of "solutions" that must be introduced when dealing with differential operators. Finally, like in any good crime story, it is only at the real end of the paper that we shall be able to revisit and compare all these examples in a unique framework.\\

\noindent
{\bf 2) MODULE THEORY}\\

Before entering the heart of this section dealing with extension modules, we need a few technical
definitions and results from commutative algebra ([17],[33],[50]).\\

\noindent
{\bf DEFINITION 2.1}: A {\it ring} $A$ is a non-empty set with two associative 
binary operations called {\it addition} and {\it multiplication},
respectively sending $a,b\in A$ to $a+b\in A$ and $ab\in A$ in such a way
that $A$ becomes an abelian group for the multiplication, so that $A$ has a
zero element denoted by $0$, every $a\in A$ has an additive inverse
denoted by $-a$ and the multiplication is distributive over the addition,
that is to say $a(b+c)=ab+ac, (a+b)c=ac+bc, \forall a,b,c\in A$.\\
A ring $A$ is said to be {\it unitary} if it has a (unique) element 
$1\in A$ such that $1a=a1=a, \forall a\in A$ and {\it commutative} if 
$ab=ba, \forall a,b\in A$. \\
A non-zero element $a\in A$ is called a {\it zero-divisor} if one can find 
a non-zero $b\in A$ such that $ab=0$ and a ring is called an 
{\it integral domain} if it has no zero-divisor.\\

\noindent
{\bf DEFINITION 2.2}: A ring $K$ is called a {\it field} if every non-zero element
$a\in K$ is a {\it unit}, that is one can find an element $b\in K$ such that 
$ab=1\in K$.\\

\noindent
{\bf DEFINITION 2.3}: A {\it module} $M$ over a ring $A$ or simply an 
$A$-{\it module} is a set of elements $x,y,z,...$ which is an abelian group
for an addition $(x,y)\rightarrow x+y$ with an action $A\times M\rightarrow
M:(a,x)\rightarrow ax$ satisfying:\\
$\bullet$ \hspace{1cm} $a(x+y)=ax+ay, \forall a\in A, \forall x,y\in M$\\
$\bullet$ \hspace{1cm} $a(bx)=(ab)x, \forall a,b\in A, \forall x\in M$\\
$\bullet$ \hspace{1cm} $(a+b)x=ax+bx, \forall a,b\in A, \forall x\in M$\\
$\bullet$ \hspace{2cm} $1x=x, \forall x\in M$ \\
The set of modules over a ring $A$ will be denoted by $mod(A)$. A module 
over a field is called a {\it vector space}.\\

\noindent
{\bf DEFINITION 2.4}: A map $f:M\rightarrow N$ between two $A$-modules is
called a {\it homomorphism} over $A$ if $f(x+y)=f(x)+f(y), \forall x,y\in
M$ and $f(ax)=af(x), \forall a\in A, \forall x\in M$. We successively
define:\\
$\bullet$ \hspace{2cm} $ker(f)=\{x\in M{\mid} f(x)=0\}$\\
$\bullet$ \hspace{2cm} $coim(f)=M/ker(f)$ \\
$\bullet$ \hspace{2cm} $im(f)=\{y\in N{\mid} \exists x\in M, f(x)=y\}$ \\
$\bullet$ \hspace{2cm} $coker(f)=N/im(f)$ \\
with an isomorphism $coim(f) \simeq im(f)$ induced by $f$.  \\

\noindent
{\bf DEFINITION 2.5}: We say that a chain of modules and homomorphisms is a 
{\it sequence} if the composition of two successive such homomorphisms is
zero. A sequence is said to be {\it exact} if the kernel of each map is
equal to the image of the map preceding it. An injective homomorphism is
called a {\it monomorphism}, a surjective homomorphism is called an 
{\it epimorphism} and a bijective homomorphism is called an 
{\it isomorphism}. A short exact sequence is an exact sequence made by a
monomorphism followed by an epimorphism.\\

The proof of the following proposition is left to the reader as an
exercise:\\

\noindent
{\bf PROPOSITION 2.6}: If one has a short exact sequence:
\[0\longrightarrow
M'\stackrel{f}{\longrightarrow}M\stackrel{g}{\longrightarrow}M''
\longrightarrow 0  \]
then the following conditions are equivalent:\\
$\bullet$ There exists an epimorphism $u:M\rightarrow M'$ such that 
$u\circ f=id_{M'}$.\\
$\bullet$ There exists a monomorphism $v:M''\rightarrow M$ such that 
$g\circ v=id_{M''}$.\\
$\bullet$ There are maps $u:M\rightarrow M'$
and $v: M''\rightarrow M$ such that $f\circ u+v\circ g=id_M$ and this relation provides an isomorphism $(u,g):M\rightarrow M'\oplus M''$ with inverse 
$f+v:M'\oplus M"\rightarrow M$.  \\

\noindent
{\bf DEFINITION 2.7}: In the above situation, we say that the short exact
sequence {\it splits} and $u(v)$ is called a {\it lift} for $f(g)$. The short exact sequence 
$0 \rightarrow \mathbb{Z} \rightarrow  \mathbb{Q} \rightarrow \mathbb{Q}/\mathbb{Z} \rightarrow 0$ cannot split over $\mathbb{Z}$. \\

\noindent
{\bf DEFINITION 2.8}: A left (right) {\it ideal} $\mathfrak{a}$ in a ring $A$ is a 
submodule of $A$ considered as a left (right) module over itself. When the
inclusion $\mathfrak{a}\subset A$ is strict, we say that $\mathfrak{a}$ is
a {\it proper ideal} of $A$.\\

\noindent
{\bf LEMMA 2.9}: If $\mathfrak{a}$ is an ideal in a ring $A$, the set of
elements $rad(\mathfrak{a})=\{a\in A{\mid} \exists n\in \mathbb{N}, 
a^n\in \mathfrak{a}\}$ is an ideal of $A$ containing $\mathfrak{a}$ and
called the {\it radical} of $\mathfrak{a}$. An ideal is called {\it perfect} 
or {\it radical} if it is equal to its radical. \\

\noindent
{\bf DEFINITION 2.10}: For any subset $S\subset A$, the smallest ideal
containing $S$ is called the ideal {\it generated} by $S$. An ideal generated 
by a single element is called a {\it principal ideal} and a ring is called
a {\it principal ideal ring} if any ideal is principal. The simplest
example is that of polynomial rings in one indeterminate over a field. When
$\mathfrak{a}$ and $\mathfrak{b}$ are two ideals of $A$, we shall denote by
$\mathfrak{a}+\mathfrak{b}$ ($\mathfrak{a}\mathfrak{b}$) the ideal
generated by all the sums $a+b$ (products $ab$) with 
$a\in\mathfrak{a}, b\in\mathfrak{b}$.\\

\noindent
{\bf DEFINITION 2.11}: An ideal $\mathfrak{p}$ of a ring $A$ is called a 
{\it prime ideal} if, whenever $ab\in \mathfrak{p}$ ($aAb\in \mathfrak{p}$ in
the non-commutative case) then either $a\in \mathfrak{p}$ or
$b\in\mathfrak{p}$. The set of proper prime ideals of $A$ is denoted by 
$spec(A)$ and called the {\it spectrum} of $A$.\\

\noindent
{\bf DEFINITION 2.12}: The {\it annihilator} of a module $M$ in $A$
is the ideal $ann_A(M)$ of $A$ made by all the elements $a\in A$ such that 
$ax=0, \forall x\in M$.\\

From now on, all rings considered will be unitary integral domains, that is
rings containing 1 and having no zero-divisor as we shall deal mainly with rings of partial differential operators. 
For the sake of clarity, as a few results will also be valid for modules over non-commutative rings, we
shall denote by ${}_AM_B$ a {\it bimodule} $M$ which is a left module for $A$ with
operation $(a,x)\rightarrow ax$ and a right module for $B$ with operation
$(x,b)\rightarrow xb$. In the commutative case, lower indices are not
needed. If $M={}_AM$ and $N={}_AN$ are two left $A$-modules, the set of
$A$-linear maps $f:M\rightarrow N$ will be denoted by $hom_A(M,N)$ or
simply $hom(M,N)$ when there will be no confusion and there is a canonical
isomorphism $hom(A,M)\simeq M:f\rightarrow f(1)$ with inverse $x\rightarrow
(a\rightarrow ax)$. When $A$ is commutative, $hom(M,N)$ is again an
$A$-module for the law $(bf)(x)=f(bx)$ as we have indeed:
\[ (bf)(ax)=f(bax)=f(abx)=af(bx)=a(bf)(x).\]
In the non-commutative case, things are much more complicate and we have:\\

\noindent
{\bf LEMMA 2.13}: Given ${}_AM$ and ${}_AN_B$, then $hom_A(M,N)$ becomes a right module over $B$ for the law $(fb)(x)=f(x)b$.\\

\noindent
{\it Proof}: We just need to check the two relations:
\[ (fb)(ax)=f(ax)b=af(x)b=a(fb)(x),\]
\[ ((fb')b")(x)=(fb')(x)b"=(f(x)b')b"=f(x)(b'b'')=(f(b'b''))(x).\]
\hspace*{12cm}               Q.E.D. \\

A similar result can be obtained with ${ }_AM_B$ and ${ }_AN$, 
where $hom_A(M,N)$ now becomes a left module over $B$ for the law $(bf)(x)=f(xb)$.\\

\noindent
{\bf THEOREM 2.14}: If $M,M',M''$ are $A$-modules, the sequence:
\[ M'\stackrel{f}{\rightarrow}M\stackrel{g}{\rightarrow} M''\rightarrow 0 \]
is exact if and only if the sequence:
\[ 0\rightarrow hom(M'',N)\rightarrow hom(M,N)\rightarrow hom(M',N) \]
is exact for any $A$-module $N$.\\

\noindent
{\it Proof}: Let us consider homomorphisms $h:M\rightarrow N$,
$h':M'\rightarrow N$, $h'':M''\rightarrow N$ such that $h''\circ g=h$,
$h\circ f=h'$. If $h=0$, then $h''\circ g=0$ implies $h''(x'')=0, \forall
x''\in M''$ because $g$ is surjective and we can find $x\in M$ such that
$x''=g(x)$. Then $h''(x'')=h''(g(x))=h''\circ g(x)=0$. Now, if $h'=0$, we
have $h\circ f=0$ and $h$ factors through $g$ because the initial sequence
is exact. Hence there exists $h'':M''\rightarrow N$ such that $h=h''\circ
g$ and the second sequence is exact.\\
We let the reader prove the converse as an exercise.\\
\hspace*{12cm}    Q.E.D.  \\

\noindent
{\bf COROLLARY 2.15}: The short exact sequence:
\[  0\rightarrow M'\rightarrow M\rightarrow M''\rightarrow 0  \]
splits if and only if the short exact sequence:
\[ 0\rightarrow hom(M'',N)\rightarrow hom(M,N)\rightarrow
hom(M',N)\rightarrow 0 \]
is exact for any module $N$.\\

\noindent
{\bf DEFINITION 2.16}: If $M$ is a module over a ring $A$, a {\it system of
  generators} of $M$ over $A$ is a family $\{x_i\}_{i\in I}$ of elements of
  $M$ such that any element of $M$ can be written $x=\sum_{i\in I}a_ix_i$
  with only a finite number of nonzero $a_i$. An $A$-module is called {\it noetherian} if every
submodule of $M$ (and thus $M$ itself) is finitely generated.\\

One has the following standard technical result:\\

\noindent
{\bf PROPOSITION 2.17}: In a short exact sequence of modules, the central module is
noetherian if and only if the two other modules are noetherian. As a byproduct, if $A$ is a noetherian ring 
and $M$ is a finitely generated module over $A$, then $M$ is noetherian.\\

Accordingly, if $M$ is generated by $\{x_1, ... ,x_r\}$, there is an epimorphism $A^r \rightarrow M: (1,0,...,0) \rightarrow x_1, ... ,(0,...,1)\rightarrow x_r$. 
The kernel of this epimorphism is thus also finitely generated, say by $\{y_1,...,y_s\}$ and we therefore
obtain the exact sequence $A^s\rightarrow A^r\rightarrow M\rightarrow 0$
that can be extended inductively to the left. Such a property will always be assumed in the sequel.\\

\noindent
{\bf DEFINITION 2.18}: In this case, we say that $M$ is 
{\it finitely presented}.\\

We now turn to the definition and brief study of tensor products of modules
over rings that will not be necessarily commutative unless stated
explicitly.\\
Let $M=M_A$ be a right $A$-module and $N={}_AN$ be a left $A$-module. We
may introduce the free $\mathbb{Z}$-module made by finite formal linear
combinations of elements of $M\times N$ with coefficients in
$\mathbb{Z}$.\\

\noindent
{\bf DEFINITION 2.19}: The tensor product of $M$ and $N$ over $A$ is the
$\mathbb{Z}$-module $M{\otimes}_AN$ obtained by quotienting the above 
$\mathbb{Z}$-module by the submodule generated by the elements of the
form:
\[  (x+x',y)-(x,y)-(x',y), (x,y+y')-(x,y)-(x,y'), (xa,y)-(x,ay) \]
and the image of $(x,y)$ will be denoted by $x\otimes y$.\\

It follows from the definition that we have the relations:
\[  (x+x')\otimes y=x\otimes y+x'\otimes y, x\otimes (y+y')=x\otimes
y+x\otimes y', xa\otimes y=x\otimes ay \]
and there is a canonical isomorphism $M{\otimes}_AA\simeq M,
A{\otimes}_AN\simeq N$. When $A$ is commutative, we may use left modules
only and $M{\otimes}_AN$ becomes a left $A$-module.\\

\noindent
{\bf EXAMPLE 2.20}: If $A=\mathbb{Z}, M=\mathbb{Z}/2\mathbb{Z}$ and
$N=\mathbb{Z}/3\mathbb{Z}$, we have 
$(\mathbb{Z}/2\mathbb{Z}){\otimes}_{\mathbb{Z}}(\mathbb{Z}/3\mathbb{Z})=0$
because $x\otimes y=3(x\otimes y)-2(x\otimes y)=x\otimes 3y-
2x\otimes y=0-0=0$.\\

We present the technique of {\it localization} in order to introduce rings and modules of fractions. We shall
define the procedure in the non-commutative case but the reader will
discover that, in the commutative case, localization is just the formal
counterpart superseding Laplace transform. However, it is essential to notice that only
the localization technique can be applied to systems with variable coefficients. We start with a basic definition:\\

\noindent
{\bf Definition 2.21}: A subset $S$ of a ring $A$ is said to be {\it
  multiplicatively closed} if $\forall s,t\in S\Rightarrow st\in S$ and
  $1\in S$.\\

In a general way, whenever $A$ is a non-commutative ring, that is $ab\neq
ba$ when $a,b\in A$, we shall set the following definition:\\

\noindent
{\bf Definition 2.22}: By a {\it left ring of fractions} or 
{\it left localization} of a noncommutative ring $A$ with respect to a 
multiplicatively closed subset $S$ of $A$, we mean a {\it ring} denoted by 
$S^{-1}A$ with a monomorphism $A\rightarrow S^{-1}A:a \rightarrow 1^{-1}a$ or simply $a$ such that:\\
1) $s$ is invertible in $S^{-1}A$, with inverse $s^{-1}1$ or simply $s^{-1}, \forall s\in S$.\\
2) Each element of $S^{-1}A$ or {\it fraction} has the form 
$s^{-1}a$ for some $s\in S, a\in A$.\\
A {\it right ring of fractions} or {\it right localization} can be
similarly defined.\\

In actual practice, we have
to distinguish carefully $s^{-1}a$ from $as^{-1}$. We shall recover the
standard notation $a/s$ of the commutative case when two fractions $a/s$
and $b/t$ can be reduced to the same denominator $st=ts$. The following
proposition is essential and will be completed by a technical lemma that
will be used for constructing localizations.\\

\noindent
{\bf Proposition 2.23}: If there exists a left localization of $A$ with respect
to $S$, then we must have $Sa\cap As\neq \emptyset, \forall a\in A, \forall s\in S$.\\

\noindent
{\it Proof}: As $S^{-1}A$ {\it must} be a ring, the element $as^{-1}$ in $S^{-1}A$ must be of
the form $t^{-1}b$ for some $t\in S, b\in A$. Accordingly, 
$as^{-1}=t^{-1}b\Rightarrow ta=bs$ with $t\in S, b\in A$. \\
\hspace*{12cm}                     Q.E.D.  \\

\noindent
{\bf Definition 2.24}: A set $S$ satisfying this condition is called a {\it left Ore set}.\\

\noindent
{\bf Lemma 2.25}: If $S$ is a left Ore set in a ring $A$, then $As\cap At\cap
S\neq \emptyset, \forall s,t \in S$ and two fractions can be brought to the same
denominator.\\

\noindent
{\it Proof}: From the left Ore condition, we can find $u\in S$ and $a\in A$
such that $us=at\in S$. More generally, we can find $u,v\in A$ such that
$us=vt\in S$ and we successively get:
\[  (us)^{-1}(ua)=s^{-1}u^{-1}ua=s^{-1}a, \hspace{3mm} (vt)^{-1}(vb)=t^{-1}v^{-1}vb=t^{-1}b  \]
so that the two fractions $s^{-1}a$ and $t^{-1}b$ can be brought to the same denominator $us=vt$. \\
\hspace*{12cm}                          Q.E.D.  \\

We are now in position to construct the ring of fractions $S^{-1}A$ whenever
$S$ satifies the two conditions of the last proposition. For this, using
the preceding lemma, let us define an equivalence relation on $S\times A$ by
saying that $(s,a)\sim (t,b)$ if one can find $u,v\in S$ such that
$us=vt\in S$ and $ua=vb$. Such a relation is clearly reflexive and
symmetric, thus we only need to prove that it is transitive. So let
$(s_1,a_1)\sim (s_2,a_2)$ and $(s_2,a_2)\sim (s_3,a_3)$. Then we can find 
$u_1,u_2\in A$ such that $u_1s_1=u_2s_2\in S$ and $u_1a_1=u_2a_2$. Also we
can find $v_2,v_3\in A$ such that $v_2s_2=v_3s_3\in S$ and
$v_2a_é=v_3a_3$. Now, from the Ore condition, one can find $w_1,w_3\in A$
such that $w_1u_1s_1=w_3v_3s_3\in S$ and thus $w_1u_2s_2=w_3v_2s_2\in S$,
that is to say $(w_1u_2-w_3v_2)s_2=0$. As $A$ is an integral domain, we have 
$w_1u_2-w_3v_2=0\Rightarrow w_1u_2=w_3v_2 \Rightarrow w_1u_1a_1=w_1u_2a_2=w_3v_2a_2=w_3v_3a_3$ as
wished. We finally define $S^{-1}A$ to be the quotient of $S\times A$ by
the above equivalence relation with $\theta:A\rightarrow
S^{-1}A:a\rightarrow 1^{-1}a$. The sum $(s,a)+(t,b)$ will be defined to be $(us=vt,ua+vb)$ and the product
$(s,a)\times (t,b)$ will be defined to be $(s^{-1}a)(t^{-1}b)= s^{-1}(at^{-1})b=s^{-1}u^{-1}cb=(us)^{-1}(cb)$ whenever $ua=ct$ .\\

A similar approach can be used in order to define and construct modules of
fractions whenever $S$ satifies the two conditions of the last
proposition. For this we need a preliminary lemma:\\

\noindent
{\bf LEMMA 2.26}: If $S$ is a left Ore set in a ring $A$ and $M$ is a left
module over $A$, the set:
\[t_S(M)=\{x\in M{\mid} \exists s\in S, sx=0\}  \]
is a submodule of $M$ called the $S$-{\it torsion submodule} of $M$.\\

\noindent
{\it Proof}: If $x,y\in t_S(M)$, we may find $s,t\in S$ such that $sx=0,
ty=0$. Now, we can find $u,v\in A$ such that $us=vt\in S$ and we
successively get $us(x+y)=usx+vty=0\Rightarrow x+y\in t_S(M)$. Also,
$\forall a\in A$, using the Ore condition for $S$, we can find $b\in A,
t\in S$ such that $ta=bs$ and we get $tax=bsx=0\Rightarrow ax\in t_S(M)$.\\
\hspace*{12cm}     Q.E.D.   \\

\noindent
{\bf DEFINITION 2.27}: By a {\it left module of fractions} or 
{\it left localization} of $M$ with respect to $S$, we mean a left module 
$S^{-1}M$ over $S^{-1}A$ both with a homomorphism
$\theta={\theta}_S:M\rightarrow S^{-1}M:x\rightarrow 1^{-1}x=s^{-1}sx$ such that:\\
1) Each element of $S^{-1}M$ has the form $s^{-1}x$ for 
$s\in S,x\in M$.\\
2) $ker({\theta}_S)=t_S(M)$.\\

In order to construct $S^{-1}M$, we shall define an equivalence relation on
$S\times M$ by saying that $(s,x)\sim (t,y)$ if there exists $u,v\in A$
such that $us=vt\in S$ and $ux=vy$. Checking that this relation is
reflexive, symmetric and transitive can be done as before (exercise) and we
define $S^{-1}M$ to be the quotient of $S\times M$ by this equivalence 
relation. The main property of localization is expressed by the following theorem:\\

\noindent
{\bf Theorem 2.28}: If one has an exact sequence:
\[    M'\stackrel{f}{\longrightarrow} M \stackrel{g}{\longrightarrow} M''
\]
then one also has the exact sequence:
\[  S^{-1}M'\stackrel{S^{-1}f}{\longrightarrow} S^{-1}M
\stackrel{S^{-1}g}{\longrightarrow} S^{-1}M''  \]
where $S^{-1}f(s^{-1}x)=s^{-1}f(x)$.\\

\noindent
{\it Proof}: As $g\circ f=0$, we also have $S^{-1}g\circ S^{-1}f=0$ and
thus $im(S^{-1}f)\subseteq ker(S^{-1}g)$.\\
In order to prove the reverse inclusion, let $s^{-1}x\in ker(S^{-1}g)$. We
have therefore $s^{-1}g(x)=0$ in $S^{-1}M''$ and there exists $t\in S$ such
that $tg(x)=g(tx)=0$ in $M''$. As the initial sequence is exact, we can
find $x'\in M'$ such that $tx=f(x')$. Accordingly, in $S^{-1}M$ we have
$s^{-1}x=s^{-1}t^{-1}tx=(ts)^{-1}tx=(ts)^{-1}f(x')=S^{-1}f((ts)^{-1}x')$
and thus $ker(S^{-1}g)\subseteq im(S^{-1}f)$.\\
\hspace*{12cm}      Q.E.D.   \\

As a link between tensor product and localization, we notice that the multiplication 
map $S^{-1}A\times M\rightarrow S^{-1}M$ given by $(s^{-1}a,x)\rightarrow
s^{-1}ax$ induces an isomorphism $S^{-1}A{\otimes}_AM\rightarrow S^{-1}M$
of modules over $S^{-1}A$ when $S^{-1}A$ is considered as a right module over
$A$ with $(s^{-1}a)b=s^{-1}ab$ and $M$ as a left module over $A$. In particular, when $A$ is a commutative integral domain and $S=A-\{0\}$, the field 
$K=Q(A)=S^{-1}A$ is called the field of fractions of $A$ and we have the
short exact sequence:
\[   0\longrightarrow A \longrightarrow K \longrightarrow K/A
\longrightarrow 0  \]
If now $M$ is a left $A$-module, we may tensor this sequence by $M$ on the
right with $A\otimes M=M$ but we do not get in general an exact sequence. 
The defect of exactness {\it on the left} is nothing else but the 
{\it torsion submodule} $t(M)=\{x\in M{\mid} \exists 0\neq s\in A, sx=0\} 
\subseteq M$ and we have the long exact sequence:
\[ 0 \longrightarrow t(M) \longrightarrow M \longrightarrow K{\otimes}_AM 
\longrightarrow K/A{\otimes}_AM \longrightarrow 0 \]
as we may describe the central map as follows:
\[  x \longrightarrow  1\otimes x=s^{-1}s\otimes x=
s^{-1}\otimes sx  \hspace{5mm},\hspace{5mm} \forall 0\neq s\in A\]
As we saw in the Introduction, such a result allows to
understand why controllability has to do with localization which is introduced implicitly through the {\it transfer matrix} in control theory. 
In particular, a module $M$ is said to be a {\it torsion module} if $t(M)=M$ and a {\it torsion-free module} if $t(M)=0$.\\

\noindent
{\bf DEFINITION 2.29}: A module in $mod(A)$ is called a {\it free module} if it
has a {\it basis}, that is a system of generators linearly independent over
$A$. When a module $F$ is free, the number of generators in a basis, and
thus in any basis, is called the {\it rank} of $F$ over $A$ and
is denoted by $rank_A(F)$. In particular, if $F$ is free of finite rank $r$,
then $F\simeq A^r$.\\

More generally, if $M$ is any module over a ring $A$ and $F$ is a maximum
free submodule of $M$, then $M/F=T$ is a torsion module. Indeed, if $x\in
M, x\notin F$, then one can find $a\in A$ such that $ax\in F$ because,
otherwise, $F\subset\{F,x\}$ should be free submodules of $M$ with a strict 
inclusion. In that case, the {\it rank} of $M$ is by definition the rank 
of $F$ over $A$. When $A$ is commutative, one has equivalently :\\

\noindent
{\bf LEMMA 2.30}: $rk_A(M)=dim_K(K{\otimes}_AM)$.\\

\noindent
{\it Proof}: Taking the tensor product by $K$ over $A$ of the short exact
sequence $0\rightarrow F\rightarrow M \rightarrow T \rightarrow 0$, we get
an isomorphism $K{\otimes}_AF\simeq K{\otimes}_AM$ because
$K{\otimes}_AT=0$ (exercise) and the lemma follows from the definition of
the rank.\\
\hspace*{12cm}   Q.E.D.  \\

We now provide two proofs of the {\it additivity property of the rank}:  \\

\noindent
{\bf PROPOSITION 2.31}: If $0\rightarrow M'\stackrel{f}{\rightarrow}
M\stackrel{g}{\rightarrow} M''\rightarrow 0$ is a short exact sequence of 
modules over a ring $A$, then we have $rk_A(M)=rk_A(M')+rk_A(M'')$.\\

\noindent
{\it Proof 1}: In the commutative case, using a localization with respect to the multiplicatively 
closed subset $S=A-\{0\}$, this proposition is just a straight consequence
of the definition of rank and the fact that localization preserves 
exactness.\\
{\it Proof 2}: Let us consider the following diagram with exact left/right
columns and central row:\\
\[
\begin{array}{rcccccl}
  & 0 & & 0 & & 0 & \\
  & \downarrow & & \downarrow & & \downarrow & \\
0\rightarrow & F'& \rightarrow & F'\oplus F''& \rightarrow &
  F''&\rightarrow 0\\
  & \;\;\;\downarrow i'& & \;\;\downarrow i & & \;\;\;\;
\downarrow i''&  \\
0\rightarrow & M'& \stackrel{f}{\rightarrow} & M &\stackrel{g}{\rightarrow}
  & M'' &\rightarrow 0 \\
  & \;\;\;\downarrow p'& & \;\;\;\downarrow p & & \;\;\;\;\downarrow p''
&   \\
0\rightarrow & T'& \rightarrow & T &\rightarrow & T''& \rightarrow 0 \\
 & \downarrow & & \downarrow & & \downarrow &   \\
 & 0 & & 0 & & 0 &  
\end{array}    \]
where $F'(F'')$ is a maximum free submodule of $M'(M'')$ and 
$T'=M'/F'(T''=M''/F'')$ is a torsion module. Pulling back by $g$ the image
under $i''$ of a basis of $F''$, we may obtain by linearity a map 
$\sigma:F''\rightarrow M$ and we define $i=f\circ i'\circ{\pi}'+
\sigma\circ{\pi}''$ where ${\pi}':F'\oplus F''\rightarrow F'$ and 
${\pi}'':F'\oplus F''\rightarrow F''$ are the canonical projections on 
each factor of the direct sum. We have $i{\mid}_{F'}=f\circ i'$ and 
$g\circ i=g\circ \sigma\circ {\pi}''=i''\circ {\pi}''$. Hence, the
diagram is commutative and thus exact with $rk_A(F'\oplus F'')=
rk_A(F')+rk_A(F'')$ trivially. Finally, if $T'$ and $T''$ are torsion
modules, it is easy to check that $T$ is a torsion module too and $F'\oplus
F''$ is thus a maximum free submodule of $M$.\\
\hspace*{12cm}                                       Q.E.D.\\

\noindent
{\bf DEFINITION 2.32}: If $f:M\rightarrow N$ is any morphism, the {\it rank} of
$f$ will be defined to be $rk_A(f)=rk_A(im(f))$.\\

We provide a few additional properties of the rank that will be used in the
sequel. For this we shall set $M^*=hom_A(M,A)$ and, for any morphism
$f:M\rightarrow N$ we shall denote by $f^*:N^*\rightarrow M^*$ the
corresponding morphism which is such that 
$f^*(h)=h\circ f, \forall h\in hom_A(N,A)$.\\

\noindent
{\bf PROPOSITION 2.33}: When $A$ is a commutative integral domain and $M$ is a
finitely presented module over $A$, then $rk_A(M)=rk_A(M^*)$.\\

\noindent
{\it Proof}: Applying $hom_A(\bullet,A)$ to the short exact sequence in the
proof of the preceding lemma while taking into account $T^*=0$, we get a
monomorphism $0\rightarrow M^*\rightarrow F^*$ and obtain therefore 
$rk_A(M^*)\leq rk_A(F^*)$. However, as $F\simeq A^r$ with $r<\infty$ 
because $M$ is finitely generated, we get $F^*\simeq A^r$ too because 
$A^*\simeq A$. It follows that $rk_A(M^*)\leq rk_A(F^*)=rk_A(F)=rk_A(M)$
and thus $rk_A(M^*)\leq rk_A(M)$.\\
Now, if $F_1\stackrel{d_1}{\rightarrow} F_0\rightarrow M\rightarrow 0$ is a
finite presentation of $M$, applying $hom_A(\bullet,A)$ to this
presentation, we get the ker/coker exact sequence:\\
\[  0 \leftarrow N \leftarrow F_1^* \stackrel{d_1^*}{\leftarrow} F_0^*
\leftarrow M^* \leftarrow 0  \]
Applying $hom_A(\bullet,A)$ to this sequence while taking into account the
isomorphisms $F_0^{**}\simeq F_0,F_1^{**}\simeq F_1$, we get the ker/coker
exact sequence:\\
\[ 0 \rightarrow N^* \rightarrow F_1 \stackrel{d_1}{\rightarrow} F_0
\rightarrow M \rightarrow 0  \]
Counting the ranks, we obtain:\\
\[  rk_A(N)-rk_A(M^*)=rk_A(F_1^*)-rk_A(F_0^*)=rk_A(F_1)-rk_A(F_0)=
rk_A(N^*)-rk_A(M) \]
and thus:\\
\[  (rk_A(M)-rk_A(M^*))+(rk_A(N)-rk_A(N^*))=0  \]
As both two numbers in this sum are non-negative, they must be zero and we
finally get $rk_A(M)=rk_A(M^*), rk_A(N)=rk_A(N^*)$.\\
\hspace*{12cm}   Q.E.D.  \\

\noindent
{\bf COROLLARY 2.34}: Under the condition of the proposition, we have 
$rk_A(f)=rk_A(f^*)$.\\

\noindent
{\it Proof}: Introducing the $ker/coker$ exact sequence:
\[ 0 \rightarrow K \rightarrow M \stackrel{f}{\rightarrow} N 
\rightarrow Q \rightarrow 0\]
we have: $rk_A(f)+rk_A(Q)=rk_A(N)$. Applying $hom_A(\bullet,A)$ and taking
into account Theorem 3.1.14, we have the exact sequence:
\[ 0\rightarrow Q^*\rightarrow N^*\stackrel{f^*}{\rightarrow} M^*  \]
and thus : $rk_A(f^*)+rk_A(Q^*)=rk_A(N^*)$. Using the preceding
proposition, we get $rk_A(Q)=rk_A(Q^*)$ and $rk_A(N)=rk_A(N^*)$, that is to
say $rk_A(f)=rk_A(f^*)$.\\
\hspace*{12cm}   Q.E.D.  \\

\noindent
{\bf 3) HOMOLOGICAL ALGEBRA}\\

We now need a few definitions and
results from homological algebra ([32],[33],[50]). In all that follows, $A,B,C,L,M,N,R,S,T, ...$ are
modules over a ring $A$ or vector spaces over a field $k$ and the linear 
maps are making the diagrams commutative. We start recalling the well known Cramer's rule for linear systems through
the exactness of the ker/coker sequence for modules. When $f:M\rightarrow N$ is a linear map (homomorphism),
we introduce the so-called ker/coker long exact sequence:  \\
\[0\longrightarrow ker(f) \longrightarrow M
\stackrel{f}{\longrightarrow} N \longrightarrow coker(f) 
\longrightarrow 0 \]
In the case of vector spaces over a field $k$, we successively have 
$rk(f)=dim(im(f))$, $dim(ker(f))=dim(M)-rk(f)$, 
$dim(coker(f))=dim(N)-rk(f)=nb$ of
compatibility conditions, and obtain by substraction:
\[ dim(ker(f))-dim(M)+dim(N)-dim(coker(f))=0 \]
In the case of modules, we may replace the dimension by
the rank and obtain the same relations because of the additive property of
the rank. The following theorem is essential:\\

\noindent
{\bf SNAKE THEOREM 3.1}: When one has the following commutative diagram
resulting from the two central vertical short exact sequences by
exhibiting the three corresponding horizontal ker/coker exact sequences:
\[
\begin{array}{ccccccccccc}
 & & 0 & & 0 & & 0 & & & & \\
 & &\downarrow & & \downarrow & & \downarrow & & & & \\
0&\longrightarrow &K&\longrightarrow &A&\longrightarrow
&A'&\longrightarrow &Q&\longrightarrow &0\\
 & &\downarrow & &\;\;\;\downarrow \! f&
&\;\;\;\;\downarrow \! f'& &\downarrow & & \\
0&\longrightarrow &L&\longrightarrow &B&\longrightarrow
&B'&\longrightarrow &R& \longrightarrow &0 \\
 & &\downarrow & &\;\;\;\downarrow \! g & &\;\;\;\;
\downarrow \! g'& & \downarrow & & \\
0 & \longrightarrow &M& \longrightarrow &C&
\longrightarrow &C'& \longrightarrow &S& \longrightarrow
&0 \\
 & & & & \downarrow & & \downarrow & & \downarrow & & \\
 & & & &0& &0& &0& &
\end{array}
\]
then there exists a connecting map $M \longrightarrow Q$ both with a long
exact sequence:
\[0 \longrightarrow K \longrightarrow L \longrightarrow M
\longrightarrow Q \longrightarrow R \longrightarrow S
\longrightarrow 0.\]\\

\noindent
{\it Proof}: We start constructing the connecting map by using the
following succession of elements:
\[
\begin{array}{ccccccc}
 & & a & \cdots & a'& \longrightarrow &q \\
 & & \vdots & & \downarrow & & \\
 & & b & \longrightarrow &b' & & \\
 & & \downarrow & & \vdots & & \\
m& \longrightarrow &c& \cdots &0& & 
\end{array}
\]
Indeed, starting with $m\in M$, we may identify it with $c\in C$ in the
kernel of the next horizontal map. As $g$ is an epimorphism, we may find
$b\in B$ such that $c=g(b)$ and apply the next horizontal map to get
$b'\in B'$ in the kernel of $g'$ by the commutativity of the lower
square. Accordingly, there is a unique $a'\in A'$ such that
$b'=f'(a')$ and we may finally project $a'$ to $q\in Q$. The map is
well defined because, if we take another lift for $c$ in $B$, it will
differ from $b$ by the image under $f$ of a certain $a\in A$ having zero
image in $Q$ by composition. The remaining of the proof is similar. The above explicit procedure called " {\it chase} " will not be
repeated. \\
\hspace*{12cm}  Q.E.D. \\

We may now introduce {\it cohomology theory} through the following 
definition:\\

\noindent
{\bf DEFINITION 3.2}: If one has a sequence $L\stackrel{f}{\longrightarrow} M \stackrel{g}{\longrightarrow} N $, that is if $g\circ f=0$, 
then one may introduce the submodules $coboundary=B=im(f)\subseteq ker(g)=cocycle=Z\subseteq M$ and define 
the cohomology at $M$ to be the quotient $H=Z/B $.\\

\noindent
{\bf THEOREM 3.3}: The following commutative diagram where the two central 
vertical sequences are long exact sequences and the horizontal lines are
ker/coker exact sequences:
\[        \begin{array}{ccccccccccccc}
 & &0& &0& &0& & & & & & \\
 & &\downarrow & & \downarrow & & \downarrow & & & & & &
\\
0&\longrightarrow &K&\longrightarrow &A&\longrightarrow
&A'&\longrightarrow &Q&\longrightarrow &0& & \\
 & &\downarrow & &\;\;\;\downarrow \! f &
&\;\;\;\;\downarrow \! f' & &\downarrow & & & & \\
0&\longrightarrow &L&\longrightarrow&B&\longrightarrow
&B'&\longrightarrow &R&\longrightarrow &0& & \\
\cdots &\cdots &\downarrow &\cdots &\;\;\;\downarrow \!
g &\cdots &\;\;\;\;\downarrow \! g'&\cdots
&\downarrow &\cdots &\cdots &\cdots &cut \\
0&\longrightarrow &M&\longrightarrow &C&\longrightarrow
&C'&\longrightarrow &S&\longrightarrow &0& & \\
 & &\downarrow & &\;\;\;\downarrow \! h &
&\;\;\;\;\downarrow \! h'& &\downarrow & & & & \\
0&\longrightarrow &N&\longrightarrow &D&\longrightarrow
&D'&\longrightarrow &T&\longrightarrow &0& & \\
 & & & &\downarrow & &\downarrow & &\downarrow & & & & \\
 & & & &0& &0& &0 & & & & 
\end{array}   \]
induces an isomorphism between the cohomology at $M$ in the left vertical
column and the kernel of the morphism $Q\rightarrow R$ in the right
vertical column.\\

\noindent
{\it Proof}: Let us ``cut'' the preceding diagram along the dotted line. We obtain the following two
commutative and exact diagrams with $im(g)=ker(h), im(g')=ker(h')$:
\[   \begin{array}{ccccccccccc}
 & &0& &0& &0 & & & \\
 & &\downarrow & &\downarrow & &\downarrow & & & & \\
 0&\longrightarrow &K&\longrightarrow &A&\longrightarrow
&A'&\longrightarrow &Q&\longrightarrow &0 \\
 & &\downarrow & &\;\;\;\downarrow \! f &
&\;\;\;\;\downarrow \! f' & &\downarrow & & \\
0&\longrightarrow &L&\longrightarrow &B&\longrightarrow
&B'&\longrightarrow &R&\longrightarrow &0 \\
 & &\downarrow & &\;\;\;\downarrow \! g &
&\;\;\;\;\downarrow \! g' & & & & \\
0&\longrightarrow & cocycle &\longrightarrow &im (g)
&\longrightarrow &im (g') & & & & \\
 & & & &\downarrow & &\downarrow & & & & \\
 & & & &0& &0& & & & 
\end{array}    \]

\[   \begin{array}{ccccccc}
 & &0& &0& &0 \\
 & &\downarrow & &\downarrow & &\downarrow \\
0&\longrightarrow & cocycle &\longrightarrow & ker (h) &\longrightarrow &ker (h')\\
 & &\downarrow & &\downarrow & &\downarrow \\
0&\longrightarrow &M&\longrightarrow &C&\longrightarrow
&C' \\
 & &\downarrow & &\;\;\;\downarrow \! h &
&\;\;\;\;\downarrow \! h' \\
0&\longrightarrow &N&\longrightarrow &D&\longrightarrow
&D' \\
 & & & &\downarrow & &\downarrow \\
 & & & &0& &0 
\end{array}   \]
Using the snake theorem, we successively obtain the following long exact sequences: \\
\[   \begin{array}{ccccc}
\Longrightarrow &\exists &\qquad &0\longrightarrow K
\longrightarrow L \stackrel{g}{\longrightarrow}
cocycle \longrightarrow Q \longrightarrow R &\qquad \\
\Longrightarrow &\exists &\qquad & 0 \longrightarrow
coboundary \longrightarrow cocycle \longrightarrow ker \,(Q\longrightarrow R) 
\longrightarrow 0 &\qquad \\
\Longrightarrow & &\qquad & cohomology \hspace{1mm} at \, M \simeq
ker \,(Q \longrightarrow R) & 
\end{array}   \]
\hspace*{12cm}   Q.E.D.  \\

We now introduce the {\it extension modules} in an elementary manner, using
the standard notation $hom_A(M,A)=M^*$. For this, we shall use a {\it free 
resolution} of an $A$-module $M$, that is to say a long exact sequence:
\[  ...\stackrel{d_2}{\longrightarrow} F_1 \stackrel{d_1}{\longrightarrow}
F_0 \longrightarrow M \longrightarrow 0  \]
where $F_0, F_1, ... $are free modules, namely modules isomorphic to
powers of $A$ and $M=coker(d_1)=F_0/im(d_1)$. We may {\it take out} $M$ and obtain the
{\it deleted sequence}:
\[  ...\stackrel{d_2}{\longrightarrow} F_1 \stackrel{d_1}{\longrightarrow}
F_0 \longrightarrow 0  \]
which is of course no longer exact. We may 
apply the functor $hom_A(\bullet,A)$ and obtain the sequence:
\[ ...\stackrel{d_2^*}{\longleftarrow} F_1^* \stackrel{d_1^*}
{\longleftarrow} F_0^* \longleftarrow 0  \]
in order to state:\\

\noindent
{\bf DEFINITION 3.4}: We set:  \\
\[ ext^0(M)=ext^0_A(M,A)=ker(d^*_1)=M^*, \hspace{5mm} ext^i(M)=ext^i_A(M,A)=ker(d^*_{i+1})/im(d^*_i), \forall i\geq 1  \]

The extension modules have the following three main properties, the first and second only being classical ([5],[32],[50]):  \\

\noindent
{\bf PROPOSITION 3.5}: The extension modules do not depend on the resolution of $M$ chosen.  \\ 

\noindent
{\bf PROPOSITION 3.6}: If $0\rightarrow M'\rightarrow M \rightarrow
M''\rightarrow 0$ is a short exact sequence of $A$-modules, then we have
the following {\it connecting long exact sequence}:
\[   0\rightarrow M"^* \rightarrow M^*\rightarrow M'^* \rightarrow ext^1(M'') \rightarrow ext^1(M) \rightarrow ext^1(M') \rightarrow ext^2(M") \rightarrow ext^2(M) \rightarrow ...   \]
of extension modules.\\

We provide two different proofs of the following proposition:\\

\noindent
{\bf PROPOSITION 3.7}: $ext^i(M)$ is a torsion module, $\forall i\geq 1$.\\

\noindent
{\it Proof} 1: Let $F$ be a maximal free submodule of $M$. From the short exact sequence: 
\[0\longrightarrow F\longrightarrow M\longrightarrow M/F\longrightarrow 0\]
where $M/F$ is a torsion module, we obtain the long exact sequence:
\[     ...\rightarrow ext^{i-1}(F)\rightarrow ext^i(M/F)\rightarrow ext^i(M)\rightarrow ext^i(F)\rightarrow ...     \]  
As $F$ is free, we obtain $ext^i(F)=0, \forall i\geq 1$ and thus $ext^i(M)\simeq
ext^i(M/F), \forall i\geq 2$. Now, we have seen that the tensor product by $K$ of any exact sequence is again an exact sequence. Accordingly, 
as $K{\otimes}_A(M/F)=0$, we have from the definitions:
\[    K{\otimes}_A ext^i_A(M/F,A)\simeq ext^i_A(M/F,K)\simeq ext^i_K(K{\otimes}_AM/F,K)=0 , \forall i\geq 1   \]
and we finally obtain from the above sequence $ K{\otimes}_Aext^i(M)=0 
\Rightarrow ext^i(M)$ torsion, $\forall i\geq 1$.\\

\noindent
{\it Proof} 2: Having in mind that $B_i=im(d_i^*)$ and
$Z_i=ker(d_{i+1}^*)$, we obtain $rk(B_i)=rk(d_i^*)=rk(d_i)$ and
$rk(Z_i)=rk(F_i^*)-rk(d_{i+1}^*)=rk(F_i)-rk(d_{i+1})$. However, we started
from a resolution, that is an exact sequence in which
$rk(d_i)+rk(d_{i+1})=rk(F_i)$. It follows that $rk(B_i)=rk(Z_i)$ and thus
$rk(H_i)=rk(Z_i)-rk(B_i)=0$, that is to say $ext^i(M)$ is a torsion
module for $i\geq 1$, $\forall M\in mod(A)$.\\
  \hspace*{12cm}  Q.E.D.   \\

The next theorem and its corollary constitute the main results that will be used for applications through a classification of modules ([4],[15],[32],[33],[40],[45],[46]):\\

\noindent
{\bf THEOREM 3.8}: The following long exact sequence: \\
\[  0 \longrightarrow ext^1(N) \longrightarrow M \stackrel{\epsilon}{\longrightarrow} M^{**} \longrightarrow ext^2(N) \longrightarrow 0  \]
is isomorphic to the {\it ker/coker} long exact sequence for the central morphism $\epsilon$ which is defined by $\epsilon (x)(f)=f(x), \forall x\in M, \forall f\in M^*$.  \\

\noindent
{\it Proof}: Introducing $K=im(d_1^*)$, we may obtain two short exact sequences, a left one starting with $K$ and a right one finishing with $K$ as follows:\\

\[  \begin{array}{rcccccccccl}
 0\leftarrow & N & \longleftarrow & F_1^*& &\stackrel{d_1^*}{ \longleftarrow }&  & F_0^* & \longleftarrow & M^* & \leftarrow 0  \\
                     &      &                          &           & \nwarrow & & \swarrow & & & &   \\
                     &    &                            &      &            & K & & & & &   \\
                     &     &       &                           & \swarrow & & \nwarrow & & & &   \\
                                                  & & & 0 & & & & 0 & & & 
  \end{array}  \]
  Using the two corresponding long exact connecting sequences, we get $ext^1(K)\simeq ext^2(N)$ from the one starting with $N^*$ which is also providing the left exact column of the next diagram and the exact central row of this diagram from the one starting with $K^*$. The Theorem is finally obtained by a chase proving that the full diagram is commutative and exact: \\
\[  \begin{array}{rcccccccl}
  & F_1 & = & F_1 & & &  & & \\
  &\downarrow & &\hspace{3mm} \downarrow d_1& & & & & \\
  0 \rightarrow & K^* & \stackrel{d_1}{\longrightarrow} & F_0 &  \longrightarrow & M^{**} & \longrightarrow & ext^2(N) & \rightarrow 0 \\
                          &  \downarrow & &\downarrow  & \nearrow & & & &  \\
  0 \rightarrow & ext^1(N) & \longrightarrow & M & & & & & \\
                          & \downarrow & & \downarrow & & & & & \\
                          &       0  & & 0 &  && & & 
                          \end{array}   \] 
 \hspace*{12cm} Q.E.D. \\
                          
\noindent
{\bf COROLLARY 3.9}: $t(M)=ext^{1}(N)=ker(\epsilon)$.  \\

\noindent
{\it Proof}: As $ext^1(N)\subseteq M$ is a torsion module, we have therefore $ext^1(N)\subseteq t(M)$. Now, if $x\in t(M)$, we may find $0\neq a\in A$ such that $ax=0$ and 
$\epsilon(x)(f)=f(x) \Rightarrow f(ax)=af(x)=0 \Rightarrow f(x)=0$ because $A$ is an integral domain, that is $t(M)\subseteq ker(\epsilon)=ext^1(N)$ and thus $t(M)=ext^1(N)=ker(\epsilon)$.  \\
\hspace*{12cm}  Q.E.D.  \\

\noindent
{\bf DEFINITION  3.10}: A module $M$ will be called {\it torsion-free} if $ext^1(N)=t(M)=0$ and {\it reflexive} if $ext^1(N)=0, ext^2(N)=0$. Going further on in specifying the properties of $M$ can be done but is out of the scope of this paper (See [4],[32] for more details and the Poincar\'{e} sequence below for an example) ({\it Hint}: Using the connecting long exact sequence, we get $ext^{2+r}(N)=ext^r(M^*), \forall r\geq 1$). \\

Despite all these results, a major difficulty still remains. Indeed, we have $M=coker(d_1)={ }_AM$ as a left module over $A$ but, using the bimodule structure of 
$A={ }_AA_A$ and Lemma 2.13, it follows that $M^*=ker(d_1^*)=M_A^*$ is a right module over $A$ and thus $N=coker(d_1^*)=N_A$ is also a right module over $A$. However, as we shall see, all the differential modules used through applications will be left modules over the ring of differential operators and it will therefore not be possible to use dual sequences as we did without being able to " {\it pass from left to right and vice-versa} ". For this purpose we now need many delicate results from differential geometry, in particular a way to deal with the {\it formal adjoint} of an operator as we did many times in the Introduction.\\

\noindent
{\bf 4) SYSTEM THEORY} \\

If $E$ is a vector bundle over the base manifold $X$ with projection $\pi$ and local coordinates $(x,y)=(x^i,y^k)$ projecting onto $x=(x^i)$ for $i=1,...,n$ and $k=1,...,m$, identifying a map with its graph, a (local) section $f:U\subset X \rightarrow E$ is such that $\pi\circ f =id$ on $U$ and we write $y^k=f^k(x)$ or simply $y=f(x)$. For any change of local coordinates $(x,y)\rightarrow (\bar{x}=\varphi(x),\bar{y}=A(x)y)$ on $E$, the change of section is $y=f(x)\rightarrow \bar{y}=\bar{f}(\bar{x})$ such that ${\bar{f}}^l(\varphi(x)\equiv A^l_k(x)f^k(x)$. The new vector bundle $E^*$ obtained by changing the {\it transition matrix} $A$ to its inverse $A^{-1}$ is called the {\it dual vector bundle} of $E$. Differentiating with respect to $x^i$ and using new coordinates $y^k_i$ in place of ${\partial}_if^k(x)$, we obtain ${\bar{y}}^l_r{\partial}_i{\varphi}^r(x)=A^l_k(x)y^k_i+{\partial}_iA^l_k(x)y^k$. Introducing a multi-index $\mu=({\mu}_1,...,{\mu}_n)$ with length $\mid \mu \mid={\mu}_1+...+{\mu}_n$ and prolonging the procedure up to order $q$, we may construct in this way, by patching coordinates, a vector bundle $J_q(E)$ over $X$, called the {\it jet bundle of order} $q$ with local coordinates $(x,y_q)=(x^i,y^k_{\mu})$ with $0\leq \mid\mu\mid \leq q$ and $y^k_0=y^k$. We have therefore epimorphisms ${\pi}^{q+r}_q:J_{q+r}(E)\rightarrow J_q(E), \forall q,r\geq 0$. For a later use, we shall set $\mu+1_i=({\mu}_1,...,{\mu}_{i-1},{\mu}_i+1,{\mu}_{i+1},...,{\mu}_n)$ and define the operator $j_q:E \rightarrow J_q(E):f \rightarrow j_q(f)$ on sections by the local formula $j_q(f):(x)\rightarrow({\partial}_{\mu}f^k(x)\mid 0\leq \mid\mu\mid \leq q,k=1,...,m)$. Finally, a jet coordinate $y^k_{\mu}$ is said to be of {\it class} $i$ if ${\mu}_1=...={\mu}_{i-1}=0, {\mu}_i\neq 0$. \\

\noindent
{\bf DEFINITION 4.1}:  A {\it system} of PD equations of order $q$ on $E$ is a vector subbundle $R_q\subset J_q(E)$ locally defined by a constant rank system of linear equations for the jets of order $q$ of the form $ a^{\tau\mu}_k(x)y^k_{\mu}=0$. Its {\it first prolongation} $R_{q+1}\subset J_{q+1}(E)$ will be defined by the equations $ a^{\tau\mu}_k(x)y^k_{\mu}=0, a^{\tau\mu}_k(x)y^k_{\mu+1_i}+{\partial}_ia^{\tau\mu}_k(x)y^k_{\mu}=0$ which may not provide a system of constant rank as can easily be seen for $xy_x-y=0 \Rightarrow xy_{xx}=0$ where the rank drops at $x=0$.\\

The next definition of {\it formal integrability} (FI) will be crucial for our purpose.\\

\noindent
{\bf DEFINITION 4.2}: A system $R_q$ is said to be {\it formally integrable} if the $R_{q+r}$ are vector bundles $\forall r\geq 0$ (regularity condition) and no new equation of order $q+r$ can be obtained by prolonging the given PD equations more than $r$ times, $\forall r\geq 0$ or, equivalently, we have induced epimorphisms ${\pi}^{q+r+1}_{q+r}:R_{q+r+1} \rightarrow R_{q+r}, \forall r\geq 0$ allowing to compute " {\it step by step} " formal power series solutions.\\

A formal test has been first sketched by C. Riquier in 1910 ([48]), then improved by M. Janet in 1920 ([12],[29]) and by E. Cartan in 1945 ([6]), finally rediscovered in 1965, totally independently, by B. Buchberger who introduced Gr\"{o}bner bases, using the name of his thesis advisor. However all these tentatives have been largely superseded and achieved in an intrinsic way, again totally independently of the previous approaches, by D.C. Spencer in 1965 ([29],[31],[52]). \\

\noindent
{\bf DEFINITION 4.3}: The family $g_{q+r}$ of vector spaces over $X$ defined by the purely linear equations $ a^{\tau\mu}_k(x)v^k_{\mu+\nu}=0$ for $ \mid\mu\mid= q, \mid\nu\mid =r $ is called the {\it symbol} at order $q+r$ and only depends on $g_q$.\\

The following procedure, {\it where one may have to change linearly the independent variables if necessary}, is the key towards the next definition which is intrinsic even though it must be checked in a particular coordinate system called $\delta$-{\it regular} (See [29] and [32] for more details):\\

\noindent
$\bullet$ {\it Equations of class} $n$: Solve the maximum number ${\beta}^n_q$ of equations with respect to the jets of order $q$ and class $n$. Then call $(x^1,...,x^n)$ {\it multiplicative variables}.\\
\[  - - - - - - - - - - - - - - - -  \]
$\bullet$ {\it Equations of class} $i$: Solve the maximum number of {\it remaining} equations with respect to the jets of order $q$ and class $i$. Then call $(x^1,...,x^i)$ {\it multiplicative variables} and $(x^{i+1},...,x^n)$ {\it non-multiplicative variables}.\\
\[ - - - - - - - - - - - - - - - - - \]
$\bullet$ {\it Remaining equations equations of order} $\leq q-1$: Call $(x^1,...,x^n)$ {\it non-multiplicative variables}.\\

\noindent
{\bf DEFINITION 4.4}: The above multiplicative and non-multiplicative variables can be visualized respectively by integers and dots in the corresponding {\it Janet board}. A system of PD equations is said to be {\it involutive} if its first prolongation can be achieved by prolonging its equations only with respect to the corresponding multiplicative variables. The following numbers are called {\it characters}:  \\
\[ {\alpha}^i_q=m(q+n-i-1)!/((q-1)!(n-i)!)-{\beta}^i_q , \hspace{3mm} \forall 1\leq i \leq n\hspace{3mm} \Rightarrow \hspace{3mm}{\alpha}^1_q\geq ... \geq {\alpha}^n_q \]
For an involutive system, $(y^{{\beta}^n_q +1},...,y^m)$ can be given arbitrarily.  \\

For an involutive system of order $q$ in the above {\it solved form}, we shall use to denote by $y_{pri}$ the {\it principal jet coordinates}, namely the leading terms of the solved equations in the sense of involution. Accordingly, any formal derivative of a principal jet coordinate is again a principal jet coordinate. The remaining jet coordinates will be called {\it  parametric jet coordinates} and denoted by $y_{par}$. We shall use a "trick" in order to study the parametric jet coordinates. Indeed, the symbol of $j_q$ is the zero symbol and is thus trivially involutive at any order $q$. Accordingly, if we introduce the {\it multiplicative variables} $x^1,...,x^i$ for the parametric jets of order $q$ and class $i$, the formal derivative or a parametric jet of strict order $q$ and class $i$ by one of its multiplicative variables is uniquely obtained and cannot be a principal jet of order $q+1$ which is coming from a uniquely defined principal jet of order $q$ and class $i$. We have thus obtained the following technical Proposition which is very useful in actual practice: \\
   
\noindent
{\bf PROPOSITION 4.5}: The principal and parametric jets of strict order $q$ of an involutive system of order $q$ have the same Janet board if we extend it to all the classes that may exist for both sets, in particular the respective empty classes.   \\
   
The following technical lemmas are straightforward consequences of the definition of an involutive system and allow to construct all the possible sets of principal or parametric jet coordinates when $m,n$ and $q$ are given (See [29] p 123-125 for more details). \\
   
 \noindent
 {\bf LEMMA 4.6}: If $y^k_{\mu}\in y_{pri}$ and $y^l_{\nu}\in y_{par}$ appear in the same equation of class $i$ in solved form, then $\nu$ is of class $\leq i$ and $l>k$ when $\nu$ is also of class $i$.  \\
 
 \noindent
 {\bf LEMMA 4.7}: If $y^k_{\mu}$ is a principal jet coordinate of strict order $q$, that is $\mid \mu \mid = q$ with ${\mu}_1=0, ..., {\mu}_{i-1}=0, {\mu}_i> 0$, then $\forall j>i$, $y^k_{\mu-1_i+1_j}$ is a principal jet coordinate and this notation has a meaning because ${\mu}_i>0$. \\
 
 \noindent
{\bf LEMMA 4.8}: If there exists an equation of class $i$, there exists also an equation of class $i+1$. Accordingly, the classes of the solved equations of an involutive symbol are an increasing chain of consecutive integers ending at $n$.  \\

\noindent
{\bf LEMMA 4.9}: The indices ${\mu}_i$ of the principal jet coordinates of strict order $q$ and class $i$ are an increasing chain of consecutive integers starting 
from $1$.  \\

\noindent
{\bf PROPOSITION 4.10}: Using the Janet board and the definition of involutivity, we get:  \\
\[   dim(g_{q+r})={\sum}_{i=1}^n\frac{(r+i-1)!}{r!(i-1)!}{\alpha}^i_q   \Rightarrow  dim(R_{q+r})=dim(R_{q-1})+{\sum}_{i=1}^n\frac{(r+i)!}{r!i!}{\alpha}^i_q  \]   \\

Let $T$ be the tangent vector bundle of vector fields on $X$, $T^*$ be the cotangent vector bundle of 1-forms on $X$ and ${\wedge}^sT^*$ be the vector bundle of s-forms on $X$ with usual bases $\{dx^I=dx^{i_1}\wedge ... \wedge dx^{i_s}\}$ where we have set $I=(i_1< ... <i_s)$. Also, let $S_qT^*$ be the vector bundle of symmetric q-covariant tensors. Moreover, if  $\xi,\eta\in T$ are two vector fields on $X$, we may define their {\it bracket} $[\xi,\eta]\in T$ by the local formula $([\xi,\eta])^i(x)={\xi}^r(x){\partial}_r{\eta}^i(x)-{\eta}^s(x){\partial}_s{\xi}^i(x)$ leading to the {\it Jacobi identity} $[\xi,[\eta,\zeta]]+[\eta,[\zeta,\xi]]+[\zeta,[\xi,\eta]]=0, \forall \xi,\eta,\zeta \in T$. We may finally introduce the {\it exterior derivative} $d:{\wedge}^rT^*\rightarrow {\wedge}^{r+1}T^*:\omega={\omega}_Idx^I \rightarrow d\omega={\partial}_i{\omega}_Idx^i\wedge dx^I$ with $d^2=d\circ d\equiv 0$ in the {\it Poincar\'{e} sequence}:\\
\[  {\wedge}^0T^* \stackrel{d}{\longrightarrow} {\wedge}^1T^* \stackrel{d}{\longrightarrow} {\wedge}^2T^* \stackrel{d}{\longrightarrow} ... \stackrel{d}{\longrightarrow} {\wedge}^nT^* \longrightarrow 0  \]

In a purely algebraic setting, one has ([29],[52]):  \\

\noindent
{\bf PROPOSITION 4.11}: There exists a map $\delta:{\wedge}^sT^*\otimes S_{q+1}T^*\otimes E\rightarrow {\wedge}^{s+1}T^*\otimes S_qT^*\otimes E$ which restricts to $\delta:{\wedge}^sT^*\otimes g_{q+1}\rightarrow {\wedge}^{s+1}T^*\otimes g_q$ and ${\delta}^2=\delta\circ\delta=0$.\\

{\it Proof}: Let us introduce the family of s-forms $\omega=\{ {\omega}^k_{\mu}=v^k_{\mu,I}dx^I \}$ and set $(\delta\omega)^k_{\mu}=dx^i\wedge{\omega}^k_{\mu+1_i}$. We obtain at once $({\delta}^2\omega)^k_{\mu}=dx^i\wedge dx^j\wedge{\omega}^k_{\mu+1_i+1_j}=0$.\\
\hspace*{12cm} Q.E.D.  \\

The kernel of each $\delta$ in the first case is equal to the image of the preceding $\delta$ but this may no longer be true in the restricted case and we set (See [31], p 85-88 for more details):   \\

\noindent
{\bf DEFINITION 4.12}: We denote by $B^s_{q+r}(g_q)\subseteq Z^s_{q+r}(g_q)$ and $H^s_{q+r}(g_q)=Z^s_{q+r}(g_q)/B^s_{q+r}(g_q)$ respectively the coboundary space, cocycle space and cohomology space at ${\wedge}^sT^*\otimes g_{q+r}$ of the restricted $\delta$-sequence which only depend on $g_q$ and may not be vector bundles. The symbol $g_q$ is said to be {\it s-acyclic} if $H^1_{q+r}=...=H^s_{q+r}=0, \forall r\geq 0$, {\it involutive} if it is n-acyclic and {\it finite type} if $g_{q+r}=0$ becomes trivially involutive for r large enough. For a later use, we notice that a symbol $g_q$ is involutive {\it and} of finite type if and only if $g_q=0$. Finally, $S_qT^*\otimes E$ is involutive $\forall q\geq 0$ if we set $S_0T^*\otimes E=E$. \\

\noindent
{\bf FI CRITERION 4.13}: If ${\pi}^{q+1}_q:R_{q+1} \rightarrow R_q$ is an epimorphism of vector bundles and $g_q$ is $2$-acyclic (involutive), then $R_q$ is formally integrable (involutive).  \\

\noindent
{\bf EXAMPLE 4.14}: The system $R_2$ defined by the three PD equations\\
\[     y_{33}=0, \hspace{5mm}y_{23}-y_{11}=0,\hspace{5mm} y_{22}=0  \]
is homogeneous and thus automatically formally integrable but $g_2$ is not involutive though finite type because $g_4=0$. Elementary computations of ranks of matrices shows that the $\delta$-map:\\
\[    0\rightarrow  {\wedge}^2T^*\otimes g_3  \stackrel{\delta}{\longrightarrow} {\wedge}^3T^*\otimes g_2 \rightarrow 0  \]
is a $3\times 3$ isomorphism and thus $g_3$ is 2-acyclic with $dim(g_3)=1$, a {\it crucial intrinsic} property totally absent from any "old" work and quite more easy to handle than its Koszul dual. \\

The main use of involution is to construct differential sequences that are made up by successive {\it compatibility conditions} (CC) of order one. In particular, when $R_q$ is involutive, the differential operator ${\cal{D}}:E\stackrel{j_q}{\rightarrow} J_q(E)\stackrel{\Phi}{\rightarrow} J_q(E)/R_q=F_0$ of order $q$ with space of solutions $\Theta\subset E$ is said to be {\it involutive} and one has the canonical {\it linear Janet sequence} ([31], p 144):\\
\[  0 \longrightarrow  \Theta \longrightarrow E \stackrel{\cal{D}}{\longrightarrow} F_0 \stackrel{{\cal{D}}_1}{\longrightarrow}F_1 \stackrel{{\cal{D}}_2}{\longrightarrow} ... \stackrel{{\cal{D}}_n}{\longrightarrow} F_n \longrightarrow 0   \]
where each other operator is first order involutive and generates the CC of the preceding one with the {\it Janet bundles} $F_r={\wedge}^rT^*\otimes J_q(E)/({\wedge}^rT^*\otimes R_q+\delta ({\wedge}^{r-1}T^*\otimes S_{q+1}T^*\otimes E))$. As the Janet sequence can be "cut at any place", that is can also be constructed anew from any intermediate operator, {\it the numbering of the Janet bundles has nothing to do with that of the Poincar\'{e} sequence for the exterior derivative}, contrary to what many physicists  still believe ($n=3$ with ${\cal{D}}=div$ provides the simplest example). Moreover, the fiber dimension of the Janet bundles can be computed at once inductively from the board of multiplicative and non-multiplicative variables that can be exhibited for $\cal{D}$ by working out the board for ${\cal{D}}_1$ and so on. For this, the number of rows of this new board is the number of dots appearing in the initial board while the number $nb(i)$ of dots in the column $i$ just indicates the number of CC of class $i$ for $i=1, ... ,n$ with $nb(i) < nb(j), \forall i<j$. When $R_q$ is {\it not} involutive but formally integrable and the $r$-prolongation of its symbol $g_q$ becomes $2$-acyclic, it is known that the generating CC are of order $r+1$ (See [31], Example 6, p 120 and previous Example). \\

\noindent
{\bf EXAMPLE 4.15}: ([21],$\S 38$, p 40 is providing the first intuition of formal integrability) The second order system $y_{11}=0, y_{13}-y_2=0$ is neither formally integrable nor involutive. Indeed, we get $d_3y_{11}-d_1(y_{13}-y_2)=y_{12}$ and $d_3y_{12}-d_2(y_{13}-y_2)=y_{22}$, that is to say {\it each first and second} prolongation does bring a new second order PD equation. Considering the new system $y_{22}=0, y_{12}=0, y_{13}-y_2=0, y_{11}=0$, the question is to decide whether this system is involutive or not. In such a simple situation, as there is no PD equation of class $3$, the evident permutation of coordinates $(1,2,3)\rightarrow (3,2,1)$ provides the following involutive second order system with one equation of class $3$, $2$ equations of class $2$ and $1$ equation of clas $1$:   \\
\[  \left\{  \begin{array}{lcl}
{\Phi}^4 \equiv y_{33}  & = & 0  \\
{\Phi}^3 \equiv y_{23} & = & 0  \\
{\Phi}^2 \equiv y_{22} & = &  0  \\
{\Phi}^1 \equiv y_{13}-y_2 & = &  0 
\end{array}
\right. \fbox{$\begin{array}{lll}
1 & 2 & 3 \\
1 & 2 & \bullet \\
1 & 2 & \bullet \\
1 & \bullet & \bullet
\end{array}$}  \]
We have ${\alpha}^3_2=0,{\alpha}^2_2=0,{\alpha}^1_2=2$ and the corresponding CC system is easily seen to be the following involutive first order system:  \\
\[  \left\{  \begin{array}{lcl}
{\Psi}^4 \equiv  d_3{\Phi}^3-d_2{\Phi}^4  & = & 0  \\
{\Psi}^3 \equiv  d_3{\Phi}^2-d_2{\Phi}^3  & = & 0  \\
{\Psi}^2 \equiv  d_3{\Phi}^1-d_1{\Phi}^4+{\Phi}^3 & = &  0  \\
{\Psi}^1 \equiv  d_2{\Phi}^1-d_1{\Phi} ^3+ {\Phi}^2   & = &  0 
\end{array}
\right. \fbox{$\begin{array}{lll}
1 & 2 & 3  \\
1 & 2 & 3  \\
1 & 2 & 3  \\
1 & 2 & \bullet
\end{array}$}  \]
The final CC system is the involutive first order system:   \\
\[  \left\{  \begin{array}{lcl}
\Omega \equiv  d_3{\Psi}^1-d_2{\Psi}^2+d_1{\Psi}^4-{\Psi}^3   & = & 0  
\end{array}
\right. \fbox{$\begin{array}{lll}
1 & 2 & 3 
\end{array}$}  \]
We get therefore the Janet sequence:    
\[    0 \longrightarrow  \Theta \longrightarrow 1 \longrightarrow 4 \longrightarrow 4 \longrightarrow 1  \longrightarrow  0    \]
We finally check that each ${\Phi}^1, {\Phi}^2,{\Phi}^3$ is separately differentially dependent on ${\Phi}^4$ because we have successively $d_3{\Phi}^3-d_2{\Phi}^4=0, \hspace{2mm} d_{33}{\Phi}^2-d_{22}{\Phi}^4=0,\hspace{2mm} d_{33}{\Phi}^1-d_{13}{\Phi}^4+d_2{\Phi}^4=0$, that is ${\Phi}^1, {\Phi}^2, {\Phi}^3$ become torsion elements when ${\Phi}^4=0$. Similarly, ${\Psi}^1$ is differentially dependent on ${\Psi}^2, {\Psi}^3, {\Psi}^4$, that is ${\Psi}^1$ becomes a torsion element when 
${\Psi}^2={\Psi}^3={\Psi}^4=0$. \\

\noindent
{\bf 5)  DIFFERENTIAL MODULES }  \\
    
Let $K$ be a {\it differential field}, that is a field containing $\mathbb{Q}$ with $n$ commuting {\it derivations} $\{{\partial}_1,...,{\partial}_n\}$ with ${\partial}_i{\partial}_j={\partial}_j{\partial}_i={\partial}_{ij}, \forall i,j=1,...,n$ such that ${\partial}_i(a+b)={\partial}_ia+{\partial}_ib, \hspace{2mm} {\partial}_i(ab)=({\partial}_ia)b+a{\partial}_ib, \forall a,b\in K$ and ${\partial}_i(1/a)= - (1/a^2){\partial}_ia, \forall a\in K$. Using an implicit summation on multi-indices, we may introduce the (noncommutative) {\it ring of differential operators} $D=K[d_1,...,d_n]=K[d]$ with elements $P=a^{\mu}d_{\mu}$ such that $\mid \mu\mid<\infty$ and $d_ia=ad_i+{\partial}_ia$. The highest value of ${\mid}\mu {\mid}$ with $a^{\mu}\neq 0$ is called the {\it order} of the {\it operator} $P$ and the ring $D$ with multiplication $(P,Q)\longrightarrow P\circ Q=PQ$ is filtred by the order $q$ of the operators. We have the {\it filtration} $0\subset K=D_0\subset D_1\subset  ... \subset D_q \subset ... \subset D_{\infty}=D$. Moreover, it is clear that $D$, as an algebra, is generated by $K=D_0$ and $T=D_1/D_0$ with $D_1=K\oplus T$ if we identify an element $\xi={\xi}^id_i\in T$ with the vector field $\xi={\xi}^i(x){\partial}_i$ of differential geometry, but with ${\xi}^i\in K$ now. It follows that $D={ }_DD_D$ is a {\it bimodule} over itself, being at the same time a left $D$-module ${ }_DD$ by the composition $P \longrightarrow QP$ and a right $D$-module $D_D$ by the composition $P \longrightarrow PQ$ with $D_rD_s=D_{r+s}, \forall r,s \geq 0$. \\

If we introduce {\it differential indeterminates} $y=(y^1,...,y^m)$, we may extend $d_iy^k_{\mu}=y^k_{\mu+1_i}$ to ${\Phi}^{\tau}\equiv a^{\tau\mu}_ky^k_{\mu}\stackrel{d_i}{\longrightarrow} d_i{\Phi}^{\tau}\equiv a^{\tau\mu}_ky^k_{\mu+1_i}+{\partial}_ia^{\tau\mu}_ky^k_{\mu}$ for $\tau=1,...,p$. Therefore, setting $Dy^1+...+Dy^m=Dy\simeq D^m$ and calling $I=D\Phi\subset Dy$ the {\it differential module of equations}, we obtain by residue the {\it differential module} or $D$-{\it module} $M=Dy/D\Phi$, denoting the residue of $y^k_{\mu}$ by ${\bar{y}}^k_{\mu}$ when there can be a confusion. Introducing the two free differential modules $F_0\simeq D^{m_0}, F_1\simeq D^{m_1}$, we obtain equivalently the {\it free presentation} $F_1\stackrel{d_1}{\longrightarrow} F_0 \rightarrow M \rightarrow 0$ of order $q$ when $m_0=m, m_1=p$ and $d_1={\cal{D}}=\Phi \circ j_q$. We shall moreover assume that ${\cal{D}}$ provides a {\it strict morphism}, namely that the corresponding system $R_q$ is formally integrable. It follows that $M$ can be endowed with a {\it quotient filtration} obtained from that of $D^m$ which is defined by the order of the jet coordinates $y_q$ in $D_qy$. We have therefore the {\it inductive limit} $0=M_{-1} \subseteq M_0 \subseteq M_1 \subseteq ... \subseteq M_q \subseteq ... \subseteq M_{\infty}=M$ with $d_iM_q\subseteq M_{q+1}$ but it is important to notice that $D_rD_q=D_{q+r} \Rightarrow D_rM_q= M_{q+r}, \forall q,r\geq 0 \Rightarrow M=DM_q, \forall q\geq 0$ {\it in this particular case}. It also follows from noetherian arguments and involution that $D_ rI_q=I_{q+r}, \forall r\geq 0$ though we have in general only $D_rI_s\subseteq I_{r+s}, \forall r\geq 0, \forall s<q$. As $K\subset D$, we may introduce the {\it forgetful functor} $for : mod(D) \rightarrow mod(K): { }_DM \rightarrow { }_KM$. \\

More generally, introducing the successive CC as in the preceding section while changing slightly the numbering of the respective operators, we may finally obtain the {\it free resolution} of $M$, namely the exact sequence $\hspace{5mm} ... \stackrel{d_3}{\longrightarrow} F_2  \stackrel{d_2}{\longrightarrow} F_1 \stackrel{d_1}{\longrightarrow}F_0\longrightarrow M \longrightarrow 0 $. In actual practice, {\it one must never forget that} ${\cal{D}}=\Phi \circ j_q$ {\it acts on the left on column vectors in the operator case and on the right on row vectors in the module case}. Also, with a slight abuse of language, when ${\cal{D}}=\Phi \circ j_q$ is involutive as in section 2 and thus $R_q=ker( \Phi)$ is involutive, one should say that $M$ has an {\it involutive presentation} of order $q$ or that $M_q$ is {\it involutive}. \\

\noindent
{\bf DEFINITION 5.1}: Setting $P=a^{\mu}d_{\mu}\in D  \stackrel{ad}{\longleftrightarrow} ad(P)=(-1)^{\mid\mu\mid}d_{\mu}a^{\mu}   \in D $, we have $ad(ad(P))=P$ and $ad(PQ)=ad(Q)ad(P), \forall P,Q\in D$. Such a definition can be extended to any matrix of operators by using the transposed matrix of adjoint operators and we get:  
\[ <\lambda,{\cal{D}} \xi>=<ad({\cal{D}})\lambda,\xi>+\hspace{1mm} {div}\hspace{1mm} ( ... )  \]
from integration by part, where $\lambda$ is a row vector of test functions and $<  > $ the usual contraction. We quote the useful formulas $[ad(\xi),ad(\eta)]=ad(\xi)ad(\eta)-ad(\eta)ad(\xi)= - ad([\xi, \eta]), \forall \xi, \eta \in T$ ({\it care about the minus sign}) and $rk_D({\cal{D}})=rk_D(ad({\cal{D}}))$ as 
in ([32], p 610-612).\\

\noindent
{\bf REMARK 5.2}: As can be seen from the examples of the Introduction, when ${\cal{D}}$ is involutive, then $ad({\cal{D}})$ may not be involutive. Also, in the differential framework, we may set $diff \hspace{2mm} rk ({\cal{D}})=m-{\alpha}^n_q={\beta}^n_q$. Comparing to similar concepts used in {\it differential algebra}, this number is just the maximum number of differentially independent equations to be found in the differential module $I$ of equations. Indeed, pointing out that differential indeterminates in differential algebra are nothing else than jet coordinates in differential geometry and using standard notations, we have $K\{y\}=lim_{q\rightarrow \infty}K[y_q]$. In that case, the differential ideal $I$ {\it automatically} generates a prime differential ideal $\mathfrak{p}\subset K\{y\}$ providing a {\it differential extension} $L/K$ with $L=Q(K\{y\}/\mathfrak{p})$ and {\it differential transcendence degree} $diff\hspace{1mm}trd (L/K)={\alpha}^n_q$, a result explaining the notations ([31]). Now, from the dimension formulas of $R_{q+r}$, we obtain at once $rk_D(M)={\alpha}^n_q$ and thus $rk_D({\cal{D}})=m - rk_D(M)={\beta}^n_q$ too in any free presentation of $M$ starting with ${\cal{D}}$. However, as we already said, ${\cal{D}}$ acts on the left in differential geometry but on the right in the theory of differential modules. In the case of an operator of order zero, we just recognize the fact that the rank of a matrix is eqal to the rank of the transposed matrix.\\

\noindent
{\bf PROPOSITION 5.3}: If $f\in aut(X)$ is a local diffeomorphisms on $X$, we may set $ x=f^{-1}(y)=g(y)$ and we have the {\it identity}:
\[   \frac{\partial}{\partial y^k}(\frac{1}{\Delta (g(y))} {\partial}_if^k(g(y))\equiv 0.   \]
If we have an operator $E\stackrel{\cal{D}}{\longrightarrow} F$, we obtain therefore an operator ${\wedge}^nT^*\otimes E^*\stackrel{ad(\cal{D})}{\longleftarrow} {\wedge}^nT^*\otimes F^*$. \\

Now, with operational notations, let us consider the two differential sequences:  \\
\[   \xi  \stackrel{{\cal{D}}}{\longrightarrow} \eta \stackrel{{\cal{D}}_1}{\longrightarrow} \zeta  \]
\[   \nu  \stackrel{ad({\cal{D}})}{\longleftarrow} \mu \stackrel{ad({\cal{D}}_1)}{\longleftarrow} \lambda   \]
where ${\cal{D}}_1$ generates all the CC of ${\cal{D}}$. Then ${\cal{D}}_1\circ {\cal{D}}\equiv 0 \Longleftrightarrow ad({\cal{D}})\circ ad({\cal{D}}_1)\equiv 0 $ but $ad({\cal{D}})$ may not generate all the CC of $ad({\cal{D}}_1)$. Passing to the module framework, we just recognize the definition of $ext^1_D(M)$. Now, exactly like we defined the differential module $M$ from $\cal{D}$, let us define the differential module $N$ from $ad(\cal{D})$. Then $ext^1_D(N)=t(M)$ does not depend on the presentation of $M$. \\

Having in mind that $D$ is a $K$-algebra, that $K$ is a left $D$-module with the standard action $(D,K) \longrightarrow K:(P,a) \longrightarrow P(a):(d_i,a)\longrightarrow {\partial}_ia$ and that $D$ is a bimodule over itself, {\it we have only two possible constructions leading to the following two definitions}:  \\

\noindent
{\bf DEFINITION 5.4}: We may define the {\it inverse system} $R=hom_K(M,K)$ of $M$ and set $R_q=hom_K(M_q,K)$ as the {\it inverse system of order} $q$. \\

\noindent
{\bf DEFINITION 5.5}: We may define the right differential module $M^*=hom_D(M,D)$.  \\

The first definition is leading to the {\it inverse systems} introduced by Macaulay in ([21]) (See [37],[43] for more details). As for the second, we have (See [4, p 21] and [32, p 483-495] and [51] for more details):  \\

\noindent
{\bf THEOREM 5.6}: When $M$ and $N$ are left $D$-modules, then $hom_K(M,N)$ and $M{\otimes}_KN$ are left $D$-modules. In particular $R=hom_K(M,K)$ is also a left $D$-module for the {\it Spencer operator}.  Moreover, the structures of left $D$-modules existing therefore on $M{\otimes}_AN$ and $hom_A(N,L)$ are now coherent with the {\it adjoint isomorphism} for $mod(D)$:  \\
\[   \varphi :  hom_D(M{\otimes}_AN,L) \stackrel{\simeq}{\longrightarrow} hom_D(M,hom_A(N,L)) \hspace{5mm} ,\forall L,M,N\in mod(D)    \]

\noindent
{\it Proof}:  For any $f\in hom_K(M,N)$, let us define:   \\
\[   (af)(m)=af(m)=f(am) \hspace{1cm} \forall a\in K, \forall m\in M\]
\[   (\xi f)(m)=\xi f(m)-f(\xi m)  \hspace{1cm}  \forall \xi ={\xi}^id_i\in T, \forall m\in M  \]
It is easy to check that $\xi a=a \xi+\xi (a)$ in the operator sense and that $\xi\eta -\eta\xi =[\xi,\eta]$ is the standard bracket of vector fields. We have in particular with $d$ in place of any $d_i$: \\
\[  \begin{array}{rcl}
((da)f)(m)=(d(af))(m)=d(af(m))-af(dm)&=&(\partial a)f(m)+ad(f(m))-af(dm)\\
       &=& (a(df))(m)+(\partial a)f(m)  \\
       &=& ((ad+\partial a)f)(m)
       \end{array}  \]
 For any $m\otimes n\in M{\otimes}_KN$ with arbitrary $m\in M$ and $n\in N$, we may then define:   \\
 \[      a(m\otimes n)=am\otimes n=m\otimes an\in M{\otimes}_AN  \]
 \[  \xi (m\otimes n)=\xi m\otimes n + m\otimes \xi n \in M{\otimes}_AN   \]
 and conclude similarly with:   \\
 \[  \begin{array}{rcl}
  (da)(m\otimes n)=d(a(m\otimes n)) & = & d(am\otimes n)\\
                         & = &  d(am)\otimes n+am\otimes dn  \\
                             & = & (\partial a)m\otimes n + a(dm)\otimes n + am\otimes dn  \\
                                & = & (ad+\partial a)(m\otimes n)
                                \end{array}    \]
Using $K$ in place of $N$, we finally get $(d_if)^k_{\mu}=(d_if)(y^k_{\mu})={\partial}_if^k_{\mu}-f^k_{\mu +1_i}$ that is {\it we recognize exactly the Spencer operator} 
and thus:\\
\[  (d_i(d_jf))^k_{\mu}={\partial}_{ij}f^k_{\mu}-({\partial}_if^k_{\mu+1_j}+{\partial}_jf^k_{\mu+1_i})+f^k_{\mu+1_i+1_j} \Rightarrow d_i(d_jf)=d_j(d_if)=d_{ij}f \]
In fact, $R$ is the {\it projective limit} of ${\pi}^{q+r}_q:R_{q+r}\rightarrow R_q$ in a coherent way with jet theory ([18],[19]).\\
The next result is entrelacing the two left structures that we have just provided through the formula $(g(m))(n)=f(m\otimes n)\in N$ defining the map $\varphi$ whenever 
$f\in hom_D(M{\otimes}_A N,L)$ is given and $\varphi (f)=g$. Using any $\xi\in T$, we get successively in $L$:  \\
\[  \begin{array}{rcl}
(\xi(g(m)))(n)& = & \xi((g(m))(n))-(g(m))(\xi n)  \\
                                   & = & \xi(f(m\otimes n))-f(m\otimes \xi n)    \\
                                   & = & f(\xi(m\otimes n))-f(m\otimes \xi n)  \\
                                   & = & f(\xi m\otimes n+m\otimes \xi n)-f(m\otimes \xi n)  \\
                                   & = & f(\xi m\otimes n)  \\
                                   & = & (g(\xi m))(n)
\end{array}  \]
and thus $ \xi(g(m))=g(\xi m), \forall m\in M $ or simply $\xi\circ g=g\circ \xi$.   \\
\hspace*{12cm}  Q.E.D.  \\

\noindent
{\bf COROLLARY 5.7}: If $M$ and $N$ are right $D$-modules, then $hom_K(M,N)$ is a left $D$-module. Moreover, if $M$ is a left $D$-module and $N$ is a right $D$-module, then $M{\otimes}_KN$ is a right $D$-module. \\

\noindent
{\it Proof}: If $M$ and $N$ are right $D$-modules, we just need to set $(\xi f)(m)=f(m\xi)-f(m)\xi, \forall \xi\in T, \forall m\in M$ and conclude as before. Similarly, if $M$ is a left $D$-module and $N$ is a right $D$-module, we just need to set $(m\otimes n)\xi=m\otimes n\xi - \xi m \otimes n$. \\
\hspace*{12cm}  Q.E.D.  \\
 
\noindent
{\bf REMARK 5.8}: When $M={Ê}_DM\in mod(D)$ and $N=N_D$, , then $hom_K(N,M)$ cannot be endowed with any left or right differential structure. Similarly, when $M=M_D$ and $N=N_D$, then $M{\otimes}_KN$ cannot be endowed with any left or right differential structure (See [4], p 24 for more details).  \\

As $D={ }_DD_D$ is a bimodule, then $M^*=hom_D(M,D)$ is a right $D$-module according to Lemma 2.13 and the module $N$ defined by the ker/coker sequence $0\longleftarrow N \longleftarrow F^*_1 \stackrel{{\cal{D}}^*}{\longleftarrow} F^*_0 \longleftarrow M^* \longleftarrow 0$ is thus a right module $N_D$.\\

\noindent
{\bf COROLLARY 5.9}: We have the {\it side changing} procedure $N_D \rightarrow N={ }_DN=hom_K({\wedge}^nT^*,N_D)$ with inverse $M={Ê}_¶M \rightarrow M_D={\wedge}^nT^*{\otimes}_K M$ whenever $M,N \in mod(D)$.  \\

\noindent
{\it Proof}: According to the above Theorem, we just need to prove that ${\wedge}^nT^*$ has a natural right module structure over $D$. For this, if $\alpha=adx^1\wedge ...\wedge dx^n\in T^*$ is a volume form with coefficient $a\in K$, we may set $\alpha.P=ad(P)(a)dx^1\wedge...\wedge dx^n$ when $P\in D$. As $D$ is generated by $K$ and $T$, we just need to check that the above formula has an intrinsic meaning for any $\xi={\xi}^id_i\in T$. In that case, we check at once:
\[  \alpha.\xi=-{\partial}_i(a{\xi}^i)dx^1\wedge...\wedge dx^n=-\cal{L}(\xi)\alpha \]
by introducing the Lie derivative of $\alpha$ with respect to $\xi$, along the intrinsic formula ${\cal{L}}(\xi)=i(\xi)d+di(\xi)$ where $i( )$ is the interior multiplication and $d$ is the exterior derivative of exterior forms. According to well known properties of the Lie derivative, we get :
\[\alpha.(a\xi)=(\alpha.\xi).a-\alpha.\xi(a), \hspace{5mm} \alpha.(\xi\eta-\eta\xi)=-[\cal{L}(\xi),\cal{L}(\eta)]\alpha=-\cal{L}([\xi,\eta])\alpha=\alpha.[\xi,\eta].  \]
\hspace*{12cm}  Q.E.D.  \\

Collecting all the results so far obtained, if a differential operator ${\cal{D}}$ is given in the framework of differential geometry, we may keep the same notation ${\cal{D}}$ in the framework of differential modules which are {\it left} modules over the ring $D$ of linear differential operators and apply duality, provided we use the notation ${\cal{D}}^*$ and deal with {\it right} differential modules or use the notation $ad({\cal{D}})$ and deal again with {\it left} differential modules by using the $left \leftrightarrow right$ {\it conversion} procedure.  \\

\noindent
{\bf DEFINITION 5.10}: If a differential operator $\xi \stackrel{\cal{D}}{\longrightarrow} \eta$ is given, a {\it direct problem} is to find (generating) {\it compatibility conditions} (CC) as an operator $\eta \stackrel{{\cal{D}}_1}{\longrightarrow} \zeta $ such that ${\cal{D}}\xi=\eta \Rightarrow {\cal{D}}_1\eta=0$. Conversely, given $\eta \stackrel{{\cal{D}}_1}{\longrightarrow} \zeta$, the {\it inverse problem} will be to look for $\xi \stackrel{\cal{D}}{\longrightarrow} \eta$ such that ${\cal{D}}_1$ generates the CC of ${\cal{D}}$ and we shall say that ${\cal{D}}_1$ {\it is parametrized by} ${\cal{D}}$ {\it if such an operator} ${\cal{D}}$ {\it is existing}.  \\

 \noindent
 {\bf REMARK 5.11}: Of course, solving the direct problem (Janet, Spencer) is {\it necessary} for solving the inverse problem. However, though the direct problem always has a solution, the inverse problem may not have a solution at all and the case of the Einstein operator is one of the best non-trivial PD counterexamples (Compare [33] to [54]). It is rather striking to discover that, in the case of OD operators, it took almost 50 years to understand that the possibility to solve the inverse problem was equivalent to the controllability of the corresponding control system (Compare [14] to [33]).\\
 
As $ad(ad(P))=P, \forall P \in D$, any operator is the adjoint of a certain operator and we get:  \\

\noindent
{\bf FORMAL TEST  5.12}: The {\it double duality test} needed in order to check whether $t(M)=0$ or not and to find out a parametrization if $t(M)=0$ has 5 steps which are drawn in the following diagram where $ad({\cal{D}})$ generates the CC of $ad({\cal{D}}_1)$ and ${\cal{D}}_1'$ generates the CC of ${\cal{D}}$:  \\
\[  \begin{array}{rcccccl}
 & & & & &  {\zeta}' &\hspace{15mm} 5  \\
 & & & & \stackrel{{\cal{D}}'_1}{\nearrow} &  &  \\
4 \hspace{15mm}& \xi  & \stackrel{{\cal{D}}}{\longrightarrow} &  \eta & \stackrel{{\cal{D}}_1}{\longrightarrow} & \zeta &\hspace{15mm}   1  \\
 &  &  &  &  &  &  \\
 &  &  &  &  &  &  \\
 3 \hspace{15mm}& \nu & \stackrel{ad({\cal{D}})}{\longleftarrow} & \mu & \stackrel{ad({\cal{D}}_1)}{\longleftarrow} & \lambda &\hspace{15mm} 2
  \end{array}  \]
\vspace*{3mm}

\noindent
{\bf THEOREM 5.13}: ${\cal{D}}_1$ parametrized by ${\cal{D}} \Leftrightarrow {\cal{D}}_1={\cal{D}}'_1 \Leftrightarrow t(M)=0 \Leftrightarrow ext^1(N)=0 $.  \\

\noindent
{\bf REMARK 5.14}: When an operator ${\cal{D}}_1$ can be parametrized by an operator ${\cal{D}}$, we may ask whether or not ${\cal{D}}$ can be again parametrized by an operator ${\cal{D}}_{-1}$ and so on. A good comparison can be made with hunting rifles as a few among them, called double rifles, are equipped with a double trigger mechanism, allowing to shoot again once one has already shot. In a mathematical manner, the question is to know whether the differential module defined by ${\cal{D}}_1$ is torsion-free, reflexive and so on. The main difficulty is that these intrinsic properties highly depend on the choice of the parametrizing operator. The simplest example is provided by the Poincar\'{e} sequence for $n=3$ made by the successive $grad, curl, div$ operators. Indeed, any student knows that $curl$ is parametrizing $div$ and that $grad$ is parametrizing $curl$. However, we may parametrize ${\partial}_1{\eta}^1+{\partial}_2{\eta}^2+{\partial}_3{\eta}^3=0$ by 
choosing ${\partial}_3{\xi}^1={\eta}^1,{\partial}_3{\xi}^2={\eta}^2, -{\partial}_1{\xi}^1-{\partial}_2{\xi}^2={\eta}^3$ with $2$ potentials $({\xi}^1,{\xi}^2)$ only instead of the usual $3$ potentials $({\xi}^1,{\xi}^2, {\xi}^3)$ and cannot proceed ahead as before. Other important examples will be provided in the next section dealing with applications, in particular the one involving Einstein equations when $n=4$.  \\

It remains to study a delicate question on which all the examples of the Introduction were focussing. Indeed, if a parametrization of a given system of OD or PD equations is possible, that is, equivalently, if the corresponding differential module is torsion-free, it appears that different parametrizations may exist with quite different numbers of potentials needed. Accordingly, it should be useful to know about the possibility to have upper and lower bounds for these numbers when $n>1$, particularly in elasticity theory, because when $n=1$, an OD module $M$ with $t(M)=0$ being {\it automatically} isomorphic to a free module $E$ and the number of potentials needed is equal to $rk_D(M)=rk_D(E)$. We shall use the language of differential modules in order to revisit and improve a few results already presented in ([45], Theorem 7+ Appendix).\\

\noindent
{\bf THEOREM 5.15}: Let $F_1 \stackrel{{\cal{D}}_1}{\longrightarrow} F_0 \longrightarrow M \rightarrow 0$ be a finite free presentation of the differential module $M=coker({\cal{D}}_1)$ and assume we already know that $t(M)=0$ by using the formal test. Accordingly, we have obtained the exact sequence $F_1 \stackrel{{\cal{D}}_1}{\longrightarrow} F_0 \stackrel{{\cal{D}}}{\longrightarrow} E$ of free differential modules where ${\cal{D}}$ is the parametrizing operator. Then, there exists other parametrizations  $F_1 \stackrel{{\cal{D}}_1}{\longrightarrow} F_0 \stackrel{{\cal{D}}'}{\longrightarrow} E'$ called {\it minimal parametrizations} and such that $coker({\cal{D}}')$ is a torsion module.  \\

\noindent
{\it Proof}: We first explain the reason for using the word " {\it minimal} ". Indeed, we have $rk_D(M)=rk_D(F_0) - rk_D(im({\cal{D}}_1))=rk_D({\cal{D}})\leq rk_D(E)$ but also $rk_D(M)=rk_D({\cal{D}}')=rk_D(E')$ and thus $rk_D(E')=rk_D(M)\leq rk_D(E)$ as a way to get a lower bound for the number of potentials but not to get a differential geometric framework. \\
While applying the formal test, in the operator language $ad({\cal{D}})$ is describing the (generating) CC of $ad({\cal{D}}_1)$ and we shall denote by 
$ad({\cal{D}}_{-1})$ the (generating) CC of $ad({\cal{D}})$ as we did in Example 1.3. In the module framework, going on with left differential modules, when $F$ is a free left module, we shall denote by $\tilde{F}$ the corresponding {\it converted} left differential module of the right differential module $F^*$. The reader not familiar with duality may look at the situations met in electromagnetism and elasticity in ([32], p 492-495). If $L=coker(ad({\cal{D}}_{-1}))\simeq im(ad({\cal{D}})) \subset {\tilde{F}}_0$ and ${\tilde{E}}'$ is the largest free differential submodule of $L$, then $T=L/{\tilde{E}}'$ is a torsion module and we have the following commutative and exact diagram:  \\
\[  \begin{array}{cccccl}
& & 0 & &  0 & \\
 & & \downarrow & & \downarrow   &  \\
 0 & \longrightarrow & {\tilde{E}}' & = & {\tilde{E}}' & \rightarrow 0  \\
 \downarrow & & \downarrow &\swarrow & \downarrow  &   \\
 {\tilde{E}}_{-1} & \stackrel{ad({\cal{D}}_{-1})}{\longrightarrow} & \tilde{E} & \rightarrow & L & \rightarrow 0  \\
 \parallel &  & \downarrow &  & \downarrow &  \\
 {\tilde{E}}_{-1} & \longrightarrow & {\tilde{E}}" & \rightarrow & T &  \rightarrow 0  \\
 \downarrow &  & \downarrow   & & \downarrow &   \\
 0 &  & 0   & & 0  &                                                                                                                                                                                                                                                                                                                                                                                                                                                                                          
 \end{array}   ÊÊÊ\]
where the central vertical monomorphism ${\tilde{E}}' \rightarrow \tilde{E}$ is obtained by pulling a basis of ${\tilde{E}}'$ back to $\tilde{E}$ as we did in the diagram of Proposition 2.31. As an illustration provided at the end of Example 4.15 where each ${\Phi}^1, {\Phi}^2, {\Phi}^3$ is {\it separately} differentially dependent on ${\Phi}^4$ in the first CC system, we have the commutative and exact diagram:\\
\[  \begin{array}{cccccl}
& & 0 & &  0 & \\
 & & \downarrow & & \downarrow   &  \\
 0 & \longrightarrow & D & = & D & \rightarrow 0  \\
 \downarrow & & \downarrow & \swarrow & \downarrow  &   \\
 D^4 & \longrightarrow & D^4 & \rightarrow & L & \rightarrow 0  \\
 \parallel &  & \downarrow &  & \downarrow &  \\
 D^4 & \longrightarrow & D^3 & \rightarrow & T &  \rightarrow 0  \\
 \downarrow &  & \downarrow   & & \downarrow &   \\
 0 &  & 0   & & 0  &                                                                                                                                                                                                                                                                                                                                                                                                                                                                                          
 \end{array}          \]ÊÊÊÊÊÊÊÊÊÊÊÊÊÊÊÊÊÊÊÊÊÊÊÊÊÊÊÊÊÊÊÊÊÊÊÊÊÊÊÊÊÊÊÊÊÊÊÊÊÊÊÊÊÊÊÊÊÊÊÊÊÊÊÊÊÊÊÊÊÊÊÊÊÊÊÊÊÊÊÊÊÊÊÊÊÊÊÊÊÊÊÊÊÊÊÊÊÊÊÊÊÊÊÊÊÊÊÊÊÊÊÊÊÊÊÊÊÊÊÊÊÊÊÊÊÊÊÊÊÊÊÊÊÊÊÊÊÊÊÊÊÊÊÊÊÊÊÊÊÊÊÊÊÊÊÊÊÊÊÊÊÊÊÊÊÊÊÊÊÊÊÊÊÊÊÊÊÊÊÊÊÊÊÊÊÊÊÊÊÊÊÊÊÊÊÊÊÊÊÊÊÊÊÊÊÊÊÊÊÊÊÊÊÊÊÊÊÊÊÊÊÊÊÊÊÊÊÊÊÊÊÊÊÊÊÊÊÊÊÊÊÊÊÊÊÊÊÊÊÊÊÊÊÊÊÊÊÊÊÊÊÊÊÊÊÊÊÊÊÊÊÊÊÊÊÊÊÊÊÊÊÊÊÊÊÊÊÊÊÊÊÊÊÊÊÊÊÊÊÊÊ
Coming back to the operators $ad({\cal{D}})$ and $ad({\cal{D}}_1)$, we get the following commutative and exact diagram allowing to define $ad({\cal{D}}')$ by composition:   \\
  ÊÊÊÊÊÊÊÊÊÊÊÊÊÊÊÊÊÊÊÊÊÊÊÊÊÊÊÊÊÊÊÊÊÊÊÊÊÊÊÊÊÊÊÊÊÊÊÊÊÊÊÊÊÊÊÊÊÊÊÊÊÊÊÊÊÊÊÊÊÊÊÊÊÊÊÊÊÊÊÊÊÊÊÊÊÊÊÊÊÊÊÊÊÊÊ
\[   \begin{array}{rcccccccl}
{\tilde{F}}_1 &  \stackrel{ad({\cal{D}}_1)}{\longleftarrow} & {\tilde{F}}_0 &  & \stackrel{ad({\cal{D}})}{\longleftarrow} &  & \tilde{E} & &  \\
  &  &  &\nwarrow  &   & \swarrow & &  &  \\
   & &\parallel & & L & & \uparrow & & \\
   &  &  &\swarrow  &   & \nwarrow & &  &  \\
 &  & {\tilde{F}}_0 &  & \stackrel{ad({\cal{D}}')}{\longleftarrow} &  & {\tilde{E}}' & \leftarrow & 0  \\
 & & & & & & \uparrow & \nwarrow &  \\
 & & & & & & 0 & & 0
\end{array}  \]
We have:  \\
\[   ad({\cal{D}}_1)\circ ad({\cal{D}}')  \equiv 0 \Rightarrow {\cal{D}}'\circ {\cal{D}}_1\equiv 0
 \Rightarrow ker({\cal{D}})=im({\cal{D}}_1)\subseteq ker ({\cal{D}}')\subset F_0  \]
and obtain by duality the following commutative and exact diagram:  \\
\vspace*{5mm}
 
\[  \begin{array}{rccccccl}
 &&& & & & & 0 \\
 && && & & \nearrow &  \\
& 0   && & & M & &  \\
 &\downarrow& & & \nearrow & \downarrow & \searrow  & \\
0\rightarrow &ker({\cal{D}}) & \rightarrow& F_0 & & \stackrel{\cal{D}}{\longrightarrow}  &  & E  \\
 && && & \downarrow & & 0  \\
& &&& & \downarrow &\nearrow &   \\
 &\downarrow&  & \parallel &  & M' & & \downarrow   \\
 &&& & \nearrow & \downarrow & \searrow & \\
0 \rightarrow & ker({\cal{D}}')&\rightarrow & F_0 & & \stackrel{{\cal{D}}'}{\longrightarrow}   & & E'   \\
 &&& & & \downarrow & & \\
 &&& & & 0 & &  
\end{array}  \]

However, though the upper sequence $F_1\stackrel{{\cal{D}}_1}{\longrightarrow} F_0 \longrightarrow M \rightarrow 0$ is exact by definition, the lower induced sequence 
$F_1 \stackrel{{\cal{D}}_1}{\longrightarrow} F_0 \longrightarrow M' \rightarrow 0$ may not be exact. With $rk_D=rk$ for simplicity, $t(M)=0$ and the induced epimorphism $M \rightarrow M' \rightarrow 0$, we obtain: \\
 \[  rk({\cal{D}})=rk(ad({\cal{D}}))=rk({\tilde{F}}_0)-rk(ad({\cal{D}}_1))=rk(F_0)-rk({\cal{D}}_1)=rk(M)  \]
 \[  rk({\cal{D}}')=rk(ad({\cal{D}}'))=rk({\tilde{E}}')=rk(L)=rk({\tilde{F}}_0)-rk(ad({\cal{D}}_1))=rk(F_0)-rk({\cal{D}}_1)=rk(M')  \]
 \[ \Rightarrow rk(M) = rk(M') \Rightarrow rk(ker(M\rightarrow M'))=0 \Rightarrow  ker(M\rightarrow M')\subseteq t(M)\subseteq (M) \]
 \[  \Rightarrow ker(M\rightarrow M')=0 \Rightarrow M\simeq M'\]
Accordingly, ${\cal{D}}'$ is a minimal parametrization of ${\cal{D}}_1$ contrary to ${\cal{D}}$ in general.\\
\hspace*{12cm}   Q.E.D.   \\

\noindent
{\bf 6) APPLICATIONS}  \\

\noindent
{\bf EXAMPLE 6.1}: {\it OD Control theory Revisited}\\
The following result is well known and can be found in any textbook of algebra ([17],[50],[32]):\\

\noindent
{\bf PROPOSITION 6.2}: If $A$ is a principal ideal domain, that is if any ideal in $A$ is generated by a single element, then any torsion-free module over $A$ is free.  \\

As this is just the case of the ring $D=K[d_x]$ when $n=1$, we obtain the following corollary of the preceding parametrizing Theorem, allowing to extend the Kalman test of controllability to systems with variable coefficients as we did in the Introduction (See [14],[33],[45],[46] for more details). \\

\noindent
{\bf COROLLARY 6.3}: If $n=1$ and ${\cal{D}}_1$ is {\it surjective}, then $t(M)=0$ if and only if $ad({\cal{D}}_1)$ is injective. \\

\noindent
{\it Proof}: If $t(M)=0$, then $M\simeq E$ is a free module according to the last Proposition. Also, as ${\cal{D}}_1$ is surjective, we have the following short exact sequence:  \\
\[  0 \rightarrow F_1 \stackrel{{\cal{D}}_1}{\longrightarrow} F_0 \longrightarrow E \rightarrow 0   \]
Using a basis $(1,0,...,0),(0,1,0,...,0), ...,(0,...,0,1)$, we may construct by linearity a morphism $E \rightarrow F_0$ in such a way that this short exact sequence splits according to Proposition 2.6. Applying duality and Corollary 2.15, we get the short exact sequence:  \\
\[         0 \leftarrow F_1^* \stackrel{{\cal{D}}_1^*}{\longleftarrow} F_0^* \longleftarrow E^* \leftarrow 0   \] 
It follows that ${\cal{D}}_1^*$ is surjective and the adjoint operator $ad({\cal{D}}_1)$ is injective because $N=0$. \\
Conversely, if ${\cal{D}}_1$ is injective, then $N=0\Rightarrow ext^1(N)=0\Rightarrow t(M)=0$ according to Theorem 5.13. Meanwhile, we have proved that, if $n=1$ and $t(M)=0$, it is always possible to find an injective parametrization $M\simeq E$. \\
\hspace*{12cm}    Q.E.D.  \\

\noindent
{\bf EXAMPLE 6.4}: {\it Classical Elasticity Revisited} \\
The {\it Killing operator} ${\cal{D}}: T \longrightarrow S_2T^*$ is a defined by $\xi\in T \rightarrow {\cal{D}}\xi={\cal{L}}(\xi)\omega = \Omega=2\epsilon \in S_2T^*$ wih ${\omega}_{rj}{\partial}_i{\xi}^r+{\omega}_{ir}{\partial}_j{\xi}^r+{\xi}^r{\partial}_r{\omega}_{ij}={\Omega}_{ij}=2{\epsilon}_{ij}$ where $\xi$ is the displacement vector, ${\cal{L}}(\xi)\omega$ is the Lie derivative of $\omega$ with respect to $\xi$ and $\epsilon$ is the infinitesimal deformation tensor of textbooks. It is a {\it Lie operator} because its solutions $\Theta \subset T$ satisfy $[\Theta,\Theta]\subset \Theta$. The corresponding first order {\it Killing system} $R_1\subset J_1(T)$ is {\it not} involutive because its symbol $g_1\subset T^*\otimes T$ is finite type with first prolongation $g_2=0$ and thus $rk({\cal{D}})=n$. Accordingly, as $\omega$ is a flat constant metric, the second order CC are described by an operator ${\cal{D}}_1$ coming from the linearization of the Riemann tensor obtained in a standard way by setting $\omega \rightarrow \omega + t \Omega$ with a small parameter $t \ll 1$, dividing by $t$ and taking the limit when $t \rightarrow 0$. \\

\noindent
$\bullet$ Airy parametrization of the stress equations when $n=2$ gives $rk(E')=rk(E)=1$ and we have thus $1$ potential only. By duality, working out the corresponding adjoint operators, we obtain the two exact sequences:  \\
\[   \begin{array}{rcccccl}
  &2 &  \stackrel{Killing}{\longrightarrow}&  3 & \stackrel{Riemann}{\longrightarrow} & 1 & \rightarrow 0  \\
0 \leftarrow &2 & \stackrel{Cauchy}{\longleftarrow} & 3 & \stackrel{Airy}{\longleftarrow} & 1 &  
\end{array}  \]
Accordingly, the canonical and the minimal parametrizations coincide when $n=2$. We discover that the Airy parametrization is nothing else than the formal adjoint of the Riemann CC for the deformation tensor: \\
\[  {\partial}_i{\xi}_j + {\partial}_j{\xi}_i={\Omega}_{ij} \Rightarrow {\partial}_{22}{\Omega}_{11}+{\partial}_{11}{\Omega}_{22}-2{\partial}_{12}{\Omega}_{12}=0  \] 
where the indices of the {\it displacement vector} $({\xi}^1,{\xi}^2)$ are lowered by means of the euclidean metric of ${\mathbb{R}}^2$. We do not believe this result is known in such a general framework.  \\

\noindent
$\bullet$ Beltrami parametrization of the stress equations  when $n=3$ gives $rk(E)=6$ and we have thus $6$ potentials. However, Maxwell/Morera parametrizations of the stress equations when $n=3$ both give $rk(E')=3$ and we have thus $3$ potentials only.\\
\[  \begin{array}{rccccccl}
  &3 & \stackrel{Killing}{\longrightarrow } & 6 & \stackrel{Riemann}{\longrightarrow}& 6 & \stackrel{Bianchi}{\longrightarrow} & 3\rightarrow 0  \\
0 \leftarrow & 3 & \stackrel{Cauchy}{\longleftarrow} &6 &\stackrel{Beltrami}{\longleftarrowÊ} & 6 &\longleftarrow &  3\\
 &  &  & & \stackrel{Maxwell}{\longleftarrow} & 3  &  &
\end{array}  \] 
Accordingly, the canonical parametrization has $6$ potentials while {\it any} minimal parametrization has $3$ potentials. We finally notice that the Cauchy operator is parametrized by the Beltrami operator which is {\it again} parametrized by the adjoint of the Bianchi operator obtained by linearizing the Bianchi identities existing for the Riemann tensor, {\it a property not held by any minimal parametrization}.\\

\noindent
$\bullet$ For $n=4$, we shall prove below that the {\it Einstein parametrization} of the stress equations is neither canonical nor minimal in the following diagram:  \\
\[   \begin{array}{rcccccccccl}
 &4 & \stackrel{Killing}{\longrightarrow} & 10 & \stackrel{Riemann}{\longrightarrow} & 20 & \stackrel{Bianchi}{\longrightarrow} & 20 & \longrightarrow & 6 & \rightarrow 0 \\
0 \leftarrow & 4 & \stackrel{Cauchy}{\longleftarrow} & 10 & \longleftarrow & 20 & \longleftarrow & 20 &  & & \\
  &    &      &  &  \stackrel{Einstein}{\longleftarrow} & 10 &  &  &  & & 
\end{array}   \]
obtained by using the fact that the Einstein operator, linearization of the Einstein tensor at the Minkowski metric, is self-adjoint, the $6$ terms being exchanged between themselves. It follows that the Einstein equations in vacuum cannot be parametrized as we have the following diagram of operators (See [32] and [33] for more details or [54] for a computer algebra exhibition of this result): \\
\[  \begin{array}{rcccl}
  &  &  &\stackrel{Riemann}{ }  & 20   \\
  & &  & \nearrow &    \\
 4 &  \stackrel{Killing}{\longrightarrow} & 10 & \stackrel{Einstein}{\longrightarrow} & 10  \\
  & & & &  \\
 4 & \stackrel{Cauchy}{\longleftarrow} & 10 & \stackrel{Einstein}{\longleftarrow} & 10 
\end{array}  \]

\noindent
$\bullet$ It remains therefore to compute all these numbers for an arbitrary dimension $n\geq 2$. For this, we notice that, as we have already described the morphism $\Phi$ with kernel $R_q$ and ${\rho}_1(\Phi)$ with kernel $R_{q+1}$, then ${\rho}_r(\Phi)$ with kernel $R_{q+r}$ may be obtained similarly " step by step ". Accordingly, the link between the FI of $R_q$ and the CC of $\cal{D}$ is expressed by the following diagram that may be used inductively:  \\   
 \[  \begin{array}{rcccccccl}
      & 0 & & 0 & & 0 & &   CC  &  \\
      & \downarrow & & \downarrow & & \downarrow & & &  \\
      0\rightarrow & g_{q+r} & \rightarrow & S_{q+r}T^*\otimes E & \stackrel{{\sigma}_r(\Phi)}{\longrightarrow} & S_rT^*\otimes F_0 & \rightarrow   & coker({\sigma}_r(\Phi)) & \rightarrow 0  \\
        &  \downarrow & & \downarrow & & \downarrow & & \downarrow   &  \\
 0 \rightarrow & R_{q+r} & \rightarrow & J_{q+r}(E) &\stackrel{{\rho}_r(\Phi)}{ \longrightarrow }  & J_r(F_0) & \rightarrow & coker({\rho}_r(\Phi)) & \rightarrow 0  \\
     &  \downarrow & &  \hspace{11mm} \downarrow {\pi}^{q+r}_{q+r-1} & & \hspace{8mm} \downarrow {\pi}^r_{r-1} & & \downarrow &   \\ 
 0 \rightarrow & R_{q+r-1} & \rightarrow & J_{q+r-1}(E) & \stackrel{ {\rho}_{r-1}(\Phi)}{ \longrightarrow } & J_{r-1}(F_0) & \rightarrow & coker({\rho}_{r-1}(\Phi) & \rightarrow 0  \\ 
    &  & & \downarrow & & \downarrow & & \downarrow    & \\
    & FI &  &  0  &  &  0  &  & 0  &    
    \end{array}   \]
    
The " {\it snake theorem} " ([32],[33],[50]) then provides the long exact {\it connecting sequence}:  \\   
\[ 0 \rightarrow g_{q+r} \rightarrow  R_{q+r} \rightarrow R_{q+r-1} \rightarrow  coker({\sigma}_r(\Phi)) \rightarrow coker({\rho}_r(\Phi)) \rightarrow coker({\rho}_{r-1}(\Phi))  \rightarrow 0  \]

If we use such a diagram for the first order Killing system with no zero or first order CC, we have $q=1, E=T,F_0=J_1(T)/R_1=S_2T^*$ and $R_1$ is formally integrable ($R_2$ involutive) if and only if $\omega$ has {\it constant Riemannian curvature}:  \\
\[   {\rho}^k_{l,ij}=c({\delta}^k_i{\omega}_{lj} - {\delta}^k_j{\omega}_{li} )  \]
with $c=0$ when $\omega$ is the flat Minkowski metric (12],[29],[31]). In this case, we may apply the Spencer 
$\delta$-map to the top row obtained with $r=2$ in order to get the commutative diagram:  \\    
\[  \begin{array}{rcccccccl}
   &  0 & & 0 & & 0 &  &  &   \\
   & \downarrow & & \downarrow & & \downarrow & & &  \\
0\rightarrow & g_3 & \rightarrow &  S_3T^*\otimes T & \rightarrow & S_2T^*\otimes F_0& \rightarrow & F_1 & \rightarrow 0  \\
   & \hspace{2mm}\downarrow  \delta  & & \hspace{2mm}\downarrow \delta & &\hspace{2mm} \downarrow \delta & & &  \\
0\rightarrow& T^*\otimes g_2&\rightarrow &T^*\otimes S_2T^*\otimes T & \rightarrow &T^*\otimes T^*\otimes F_0 &\rightarrow & 0 &  \\
   &\hspace{2mm} \downarrow \delta &  &\hspace{2mm} \downarrow \delta & &\hspace{2mm}\downarrow \delta &  &  &   \\
0\rightarrow & {\wedge}^2T^*\otimes g_1 & \rightarrow & \underline{{\wedge}^2T^*\otimes T^*\otimes T} & \rightarrow & {\wedge}^2T^*\otimes F_0 & \rightarrow & 0 &  \\
   &\hspace{2mm}\downarrow \delta  &  & \hspace{2mm} \downarrow \delta  &  & \downarrow  & &  &  \\
0\rightarrow & {\wedge}^3T^*\otimes T & =  & {\wedge}^3T^*\otimes T  &\rightarrow   & 0  &  &  &   \\
    &  \downarrow  &  &  \downarrow  &  &  &  &  &  \\
    &  0  &   & 0  & &  &  &  &
\end{array}  \]
with exact rows and exact columns but the first that may not be exact at ${\wedge}^2T^*\otimes g_1$. We shall denote by $B^2(g_1)$ the {\it coboundary} as the image of the central $\delta$, by $Z^2(g_1)$ the {\it cocycle} as the kernel of the lower $\delta$ and by $H^2(g_1)=Z^2(g_1)/B^2(g_1)$ the {\it Spencer} $\delta$-{\it  cohomology} at ${\wedge}^2T^*\otimes g_1$ as the quotient.  \\

In the classical Killing system, $g_1\subset T^*\otimes T$ is defined by ${\omega}_{rj}(x){\xi}^r_i+{\omega}_{ir}(x){\xi}^r_j=0 \Rightarrow {\xi}^r_r=0, g_2=0,g_3=0$. Applying the previous diagram, we discover that the {\it Riemann tensor} $({\rho}^k_{l,ij})\subset {\wedge}^2T^*\otimes T^*\otimes T$ is a section of the bundle 
$ Riemann=F_1=H^2(g_1)=Z^2(g_1)$ with $dim(Riemann)= (n^2(n+1)^2/4)-(n^2(n+1)(n+2)/6)=(n^2(n-1)^2/4)-(n^2(n-1)(n-2)/6)=n^2(n^2-1)/12$ by using the top row or the left column. We discover at once the two properties of the (linearized) Riemann tensor through the chase involved, namely $(R^k_{l,ij})\in {\wedge}^2T^*\otimes T^*\otimes T$ is killed by both $\delta$ and ${\sigma}_0(\Phi)$. However, we have no indices for $F_1$ and cannot therefore exhibit the {\it Ricci tensor} or the {\it Einstein tensor} of general relativity by means of the usual {\it contraction} or {\it trace}. We recall briefly their standard definitions by stating $R_{ij}=R_{ji}=R^r_{i,rj} \Rightarrow R={\omega}^{ij}R_{ij}\Rightarrow E_{ij}=R_{ij}-\frac{1}{2}{\omega}_{ij}R$. Similarly, going one step further, we get the (linearized) Bianchi identities with $Bianchi=F_2=H^3(g_1)=Z^3(g_1)\Rightarrow dim(Bianchi)=dim({\wedge}^4T^*\otimes T)-dim({\wedge}^3T^*\otimes g_1)=n^2(n^2-1)(n-2)/24$. This approach is relating for the first time the concept of {\it Riemann tensor candidate}, introduced by Lanczos and others, to the Spencer $\delta$-cohomology of the Killing symbols.  \\

Counting the differential ranks is now easy because $R_1$ is formally integrable with finite type symbol and thus $R_2$ is involutive with symbol $g_2=0$. We get:  \\
\[ rk(Killing)=rk(Cauchy)=n \Rightarrow rk(Riemann)=dim(S_2T^*)-n=(n(n+1)/2) -n=n(n-1)/2\]
\[ rk(Bianchi)=(n^2(n^2-1)/2)-(n(n-1)/2)=n(n-1)(n-2)(n+3)/12 \]
that is $rk(Bianchi)=3$ when $n=3$ and $rk(Bianchi)=14=20-6$ when $n=4$. Collecting all the results, we obtain that the canonical parametrization needs $n^2(n^2-1)/12$ potentials while the minimal parametrization only needs $n(n-1)/2$ potentials. The Einstein parametrization is thus " {\it in between} " because $n(n-1)/2< n(n+1)/2 < n^2(n^2-1)/12, \forall n\geq 4$.\\
   
 The {\it conformal Killing system} ${\hat{R}}_1\subset J_1(T) $ is defined by eliminating the function $A(x)$ in the system ${\cal{L}}(\xi)\omega=A(x)\omega$. It is also a {\it Lie operator} $\hat{\cal{D}}$ with solutions $\hat{\Theta}\subset T$ satisfying $[\hat{\Theta},\hat{\Theta}]\subset \hat{\Theta}$. Its symbol ${\hat{g}}_1$ is defined by the linear equations ${\omega}_{rj}{\xi}^r_i+{\omega}_{ir}{\xi}^r_j - \frac{2}{n}{\omega}_{ij}{\xi}^r_r=0$ which do not depend on any conformal factor and is finite type when $n\geq 3$ because $g_3=0$ but ${\hat{g}}_2$ is {\it now} $2$-acyclic {\it only when} $n\geq 4$ and $3$-acyclic {\it only when} $n\geq 5$ 
 ([29],[30],[44]). It is known that ${\hat{R}}_2$ and thus ${\hat{R}}_1$ too (by a chase) are formally integrable if and only if $\omega$ has zero {\it Weyl tensor}:  \\
 \[  {\tau}^k_{l,ij}\equiv {\rho}^k_{l,ij} - \frac{1}{(n-2)}({\delta}^k_i{\rho}_{lj} - {\delta}^k_j{\rho}_{li} +{\omega}_{lj}{\omega}^{ks}{\rho}_{si} - {\omega}_{li}{\omega}^{ks}{\rho}_{sj}) + \frac{1}{(n-1)(n-2)}({\delta}^k_i{\omega}_{lj} - {\delta}^k_j{\omega}_{li})\rho=0  \]
if we use the formula $id_M-f\circ u=v\circ g$ of Proposition 2.6 in the {\it split short exact sequence} ([30],[38],[41]):  \\
\[ 0 \longrightarrow Ricci \longrightarrow Riemann \longrightarrow Weyl \longrightarrow  0  \]
according to the Vessiot structure equations, in particular if $\omega$ has constant Riemannian curvature and thus ${\rho}_{ij}={\rho}^r_{i,rj}=c(n-1){\omega}_{ij} \Rightarrow \rho={\omega}^{ij}{\rho}_{ij}=cn(n-1)$ ([29],[31],[41],[42]). Using the same diagrams as before, we get $Weyl={\hat{F}}_1=H^2({\hat{g}}_1)\neq Z^2({\hat{g}}_1)$ for defining any {\it Weyl tensor candidate}. As a byproduct, the linearized Weyl operator is of order $2$ with a symbol ${\hat{h}}_2\subset S_2T^*\otimes {\hat{F}}_0$ which is {\it not} $2$-acyclic by applying the $\delta$-map to the short exact sequence:  \\
 \[  0 \rightarrow {\hat{g}}_{3+r} \longrightarrow S_{3+r}T^*\otimes T \stackrel{{\sigma}_{2+r}(\Phi)}{\longrightarrow} {\hat{h}}_{2+r} \rightarrow  0  \]
and chasing through the commutative diagram thus obtained with $r=0,1,2$. As ${\hat{h}}_3$ becomes $2$-acyclic after one prolongation of ${\hat{h}}_2$ only, it follows that {\it the generating CC for the Weyl operator are of order} $2$, a result showing that the so-called Bianchi identities for the Weyl tensor are {\it not} CC in the strict sense of the definition as they do not involve only the Weyl tensor. Of course, these results could not have been discovered by Lanczos and followers because the formal theory of Lie pseudogroups and the Vessiot structure equations are still not acknowledged today.  \\ 

For this reason, we provide a few hints in order to explain why the Vessiot structure equations {\it sometimes contain a few constants, sometimes none at all} as we just saw (See [30],[31] and [39] for more details). Isometries preserve the metric $\omega=({\omega}_{ij}={\omega}_{ji})\in S_2T^*$ while conformal isometries preserve the symmetric tensor density $\hat{\omega}=({\hat{\omega}}_{ij}={\omega}_{ij}/({\mid det(\omega)\mid}^\frac{1}{n}))$. The respective variations are related by the similitude formula ${\hat{\Omega}}_{ij}\sim {\Omega}_{ij} - \frac{1}{n}{\omega}_{ij}{\omega}^{rs}{\Omega}_{rs}$ which only depends on $\omega$ and not on a conformal factor. It follows that $F_0=S_2T^*$ and that ${\hat{F}}_0$ may be identified with the sub-bundle $\{ \hat{\Omega}\in S_2T^*\mid {\omega}^{ij}{\hat{\Omega}}_{ij}=0\}$ with the above well defined epimorphism $F_0 \rightarrow {\hat{F}}_0$ induced by the inclusion $R_1 \subset {\hat{R}}_1$. We set ([30],[31],[39]):  \\

\noindent
{\bf DEFINITION 6.5}: We say that a vector bundle $F$ is associated with a Lie operator ${\cal{D}}$ if, for any solution $\xi\in \Theta \subset T$ of ${\cal{D}}$, there exists a first order operator 
${\cal{L}}(\xi):F \rightarrow F$ called {\it Lie derivative} with respect to $\xi$ and such that:\\
1) ${\cal{L}}(\xi + \eta)={\cal{L}}(\xi) + {\cal{L}}(\eta)    \hspace{1cm} \forall \xi,\eta \in \Theta$ \\
2) $[{\cal{L}}(\xi),{\cal{L}}(\eta)]={\cal{L}}([\xi,\eta])  \hspace{1cm}  \forall \xi,\eta \in \Theta  $\\
3) ${\cal{L}}(\xi)(f \eta)=f {\cal{L}}(\xi)\eta + (\xi . f)\eta  \hspace{1cm}  \forall \xi\in \Theta,\forall  f\in C^{\infty}(X),\forall \eta \in F$  \\
4) If $E$ and $F$ are two such associated vector bundles, then:  \\
\[{\cal{L}}(\xi)(\eta\otimes \zeta)={\cal{L}}(\xi)\eta\otimes \zeta+\eta\otimes {\cal{L}}(\xi)\zeta,\hspace{2mm} \forall \xi\in \Theta, \forall \eta\in E, \forall \zeta\in F\]

In such a case, we may introduce $ \Upsilon =\Upsilon(F)=\{\eta \in F \mid {\cal{L}}(\xi)\eta=0, \hspace{3mm}\forall \xi \in \Theta\subset T\} $.  \\
  
\noindent
{\bf PROPOSITION 6.6}: Using capital letters for linearized objects, we have:  \\
1) $\Upsilon(T)=C(\Theta)=\{\eta\in T\mid [\xi,\eta]=0, \forall \xi\in \Theta\}=centralizer$ of $\Theta$ in $T$.  \\
2) ${\Upsilon}_0=\Upsilon(F_0)=\Upsilon(S_2T^*)=\{ \Omega=A\omega\in S_2T^* \mid A=cst \}$.  \\
3) ${\Upsilon}_1=\Upsilon(F_1)=\{ R^k_{l,ij}=C({\delta}^k_i{\omega}_{lj}-{\delta}^k_j{\omega}_{li})\in F_1 \mid C=cst \}$.  \\
4) ${\hat{\Upsilon}} _1=\Upsilon({\hat{F}}_1)=0 \hspace{2mm}\Rightarrow \hspace{2mm}\Upsilon(Ricci)\simeq \Upsilon(Riemann)$.  \\     
5) The Lie derivative commutes with the Janet operators ${\cal{D}},{\cal{D}}_1,...,{\cal{D}}_n$.  \\
We have in particular ${\cal{D}}_1:{\Upsilon}_0\rightarrow {\Upsilon}_1: A\rightarrow C= - cA$ ({\it care to sign}).  \\   

\noindent
{\it Proof}: Two (nondegenerate) metrics $\omega,\bar{\omega}\in S_2T^*$ give the same Killing system $R_1$ if and only if $\bar{\omega}=a\omega$ with 
the multiplicative group parameter $a=cst$. Therefore, if $R_1$ is FI, then the two metrics have respective constant curvatures $c$ and $\bar{c}=c/a$. Setting 
$a=1+tA+...\Rightarrow \bar{c}=c+tC+...$ while linearizing these finite transformations with $t\ll 1$ gives $C=-cA$ when $t\rightarrow 0$.\\
\hspace*{12cm}  Q.E.D.  \\

However, we have yet not proved the most difficult result that could not be obtained without homological algebra and the next example will explain this additional difficulty.\\

\noindent
{\bf EXAMPLE 6.7}:  With ${\partial}_{22}\xi={\eta}^2, {\partial}_{12}\xi={\eta}^1$ for $\cal{D}$, we get  ${\partial}_1{\eta}^2-{\partial}_2{\eta}^1=\zeta$ for ${\cal{D}}_1$. Then $ad({\cal{D}}_1)$ is defined by ${\mu}^2=-{\partial}_1\lambda, {\mu}^1={\partial}_2\lambda$ while $ad(\cal{D})$ is defined by $\nu={\partial}_{12}{\mu}^1+{\partial}_{22}{\mu}^2$ but the CC of $ad({\cal{D}}_1)$ are generated by ${\nu}'={\partial}_1{\mu}^1+{\partial}_2{\mu}^2$. In the operator framework, we have the differential sequences:\\  
\[  \begin{array}{rcccl}
 \xi & \stackrel{\cal{D}}{\longrightarrow} & \eta & \stackrel{{\cal{D}}_1}{\longrightarrow} & \zeta   \\
  \nu& \stackrel{ad(\cal{D})}{\longleftarrow} & \mu & \stackrel{ad({\cal{D}}_1)}{\longleftarrow} & \lambda 
  \end{array}  \]
where the upper sequence is formally exact at $\eta$ but the lower sequence is not formally exact at $\mu$.  \\
Passing to the module framework, we obtain the sequences:  \\
\[  \begin{array}{rccccl}
 D & \stackrel{{\cal{D}}_1}{\longrightarrow} & D^2 & \stackrel{\cal{D}}{\longrightarrow} & D & \longrightarrow M \longrightarrow  0  \\
  D& \stackrel{ad({\cal{D}}_1)}{\longleftarrow} & D^2 & \stackrel{ad(\cal{D})}{\longleftarrow} & D &  
  \end{array}  \]
where the lower sequence is not exact at $D^2$.  \\

Therefore, we have to prove that the extension modules vanish, that is $ad({\cal{D}})$ generates the CC of $ad({\cal{D}}_1)$ and, conversely, that ${\cal{D}}_1$ generates the CC of ${\cal{D}}$. We also remind the reader that it has not been easy to exhibit the CC of the Maxwell or Morera parametrizations when $n=3$ and that a direct checking for $n=4$ should be strictly impossible. It has been proved by L. P. Eisenhart in 1926 ([12]) that the solution space $\Theta$ of the Killing system has $n(n+1)/2$ {\it infinitesimal generators} $\{ {\theta}_{\tau}\}$ linearly independent over the constants if and only if $\omega $ had constant Riemannian curvature, namely zero in our case. As we have a Lie group of transformations preserving the metric, the three theorems of Sophus Lie assert than $[{\theta}_{\rho},{\theta}_{\sigma}]=c^{\tau}_{\rho \sigma} {\theta}_{\tau}$ where the {\it structure constants} $c$ define a Lie algebra ${\cal{G}}$. We have therefore $\xi \in \Theta \Leftrightarrow \xi = {\lambda}^{\tau}{\theta}_{\tau}$ with ${\lambda}^{\tau}=cst$. Hence, we may replace the Killing system by the system ${\partial}_i{\lambda}^{\tau}=0$, getting therefore the differential sequence:  \\
\[  0 \rightarrow \Theta \rightarrow {\wedge}^0T^*\otimes {\cal{G}} \stackrel{d}{\longrightarrow} {\wedge}^1T^*\otimes {\cal{G}} \stackrel{d}{\longrightarrow} ... \stackrel{d}{\longrightarrow} {\wedge}^nT^* \otimes {\cal{G}} \rightarrow 0  \]
 which is the tensor product of the Poincar\'{e} sequence by ${\cal{G}}$. Finally, it follows from Proposition 3.5 that the extension modules do not depend on the resolution used and thus vanish because the Poincar\'{e} sequence is self adjoint (up to sign), that is $ad(d)$ generates the CC of $ad(d)$ at any position, exactly like $d$ generates the CC of $d$ at any position. This (difficult) result explains why the differential modules we have met were torsion-free, reflexive, ... and so on. We invite the reader to compare with the situation of the Maxwell equations in electromagnetisme. However, we have proved in ([32],[36],[41],[42],[44]) why neither the Janet sequence nor the Poincar\'{e} sequence can be used in physics and must be replaced by another resolution of $\Theta$ called {\it Spencer sequence} ([16]). \\

\noindent
{\bf EXAMPLE 6.8}: {\it PD Control Theory Revisited}  \\
Comparing with the Theorem allowing to construct a minimal parametrization, we started with ${\cal{D}}_1\eta=\zeta$ and computed $ad({\cal{D}}_1)\lambda=\mu$ with generating CC $ad({\cal{D}})\mu=\nu$, obtaining therefore finally the generating CC $ad({\cal{D}}_{-1})\nu=0$, that is ${\partial}_2{\nu}^2+{\partial}_1{\nu}^1+x^2{\nu}^1=0$. In that case, the key diagram providing the minimal parametrization is:  \\
\[  \begin{array}{cccccl}
& & 0 & &  0 & \\
 & & \downarrow & & \downarrow   &  \\
 0 & \longrightarrow & D & = & D & \rightarrow 0  \\
 \downarrow & & \downarrow & \swarrow & \downarrow  &   \\
 D & \stackrel{ad({\cal{D}}_{-1})}{\longrightarrow} & D^2 & \rightarrow & L & \rightarrow 0  \\
 \parallel &  & \downarrow &  & \downarrow &  \\
 D & \longrightarrow & D & \rightarrow & T &  \rightarrow 0  \\
 \downarrow &  & \downarrow   & & \downarrow &   \\
 0 &  & 0   & & 0  &                                                                                                                                                                                                                                                                                                                                                                                                                                                                                          
 \end{array}          \]ÊÊÊÊÊÊÊÊÊÊÊÊÊÊÊÊÊÊÊÊ
 
 This result explains why we had the potentials $({\xi}^1,{\xi}^2)$ in the canonical parametrization and $({\xi}^1=\xi,0)$ or $(0,{\xi}^2={\xi}')$ in the two minimal parametrizations exhibited. We do not believe it is possible to imagine the underlying procedure, even on such a simple example. \\  
 
 \noindent
 {\bf EXAMPLE 6.9}: {\it OD/PD Optimal Control Revisited}  \\
 Using the notations of the Formal Test 5.12, let us assume that the two differential sequences:  \\
\[  \begin{array}{rcccl}
 \xi & \stackrel{\cal{D}}{\longrightarrow} & \eta & \stackrel{{\cal{D}}_1}{\longrightarrow} & \zeta   \\
  \nu& \stackrel{ad(\cal{D})}{\longleftarrow} & \mu & \stackrel{ad({\cal{D}}_1)}{\longleftarrow} & \lambda 
  \end{array}  \] 
are {\it formally exact}, that is ${\cal{D}}_1$ generates the CC of ${\cal{D}}$ {\it and} $ad({\cal{D}})$ generates the CC of $ad({\cal{D}}_1)$, namely $\xi$ is a potential for ${\cal{D}}_1$ {\it and} $\lambda$ is a potential for $ad({\cal{D}})$. We may consider a variational problem for a cost function $\varphi (\eta)$ under the linear OD or PD constraint described by ${\cal{D}}_1\eta=0$. \\
$\bullet$ Introducing convenient Lagrange multipliers $\lambda$ while setting $dx=dx^1\wedge ... \wedge dx^n$ for simplicity, we must vary the integral:  \\
 \[       \Phi=\int [\varphi(\eta) +\lambda {\cal{D}}_1\eta]dx  \Rightarrow \delta \Phi=\int [(\partial\varphi(\eta)/\partial\eta)\delta\eta+\lambda{\cal{D}}_1\delta\eta]dx \]
Integrating by parts, we obtain the EL equations: \\
\[  \partial\varphi(\eta)/\partial\eta + ad({\cal{D}}_1)\lambda=0  \]
to which we have to add the constraint ${\cal{D}}_1\eta=0$ obtained by varying $\lambda$. If $ad({\cal{D}}_1)$ is an injective operator, in particular if ${\cal{D}}_1$ is formally surjective (no CC) and parametrized by ${\cal{D}}$, then one can obtain $\lambda$ {\it explicitly} and eliminate it by substitution. Otherwise, using the CC $ad({\cal{D}})$ of $ad({\cal{D}}_1)$, we have to study the formal integrability of the combined system: \\
\[     ad({\cal{D}})\partial\varphi(\eta)/\partial\eta=0, \hspace{4mm} {\cal{D}}_1\eta=0   \]
which may be a difficult task as we already saw through the examples of the Introduction.\\
$\bullet$ {\it We may also transform the given variational problem with constraint into a variational problem without any constraint if and only if the differential constraint can be parametrized}. Using the parametrization of ${\cal{D}}_1$ by ${\cal{D}}$, we may vary the integral: \\
\[  \Phi=\int \varphi({\cal{D}}\xi)dx  \Rightarrow \delta \Phi = \int (\partial\varphi(\eta)/\partial\eta){\cal{D}}\delta\xi  dx                \]
whenever $\eta={\cal{D}}\xi$ and integrate by parts for arbitrary $\delta\xi$ in order to obtain the EL equations:  \\
\[   ad({\cal{D}})\partial\varphi(\eta)/\partial\eta=0, \hspace{4mm}  \eta={\cal{D}}\xi  \] 
 in a coherent way with the previous approach and Poincar\'{e} duality $geometry \leftrightarrow physics$. \\
 As a byproduct, if the {\it field equations} ${\cal{D}}_1\eta=0$ can be parametrized by a {\it potential} $\xi$ through the formula ${\cal{D}}\xi=\eta$, then the {\it induction equations} $ad({\cal{D}})\mu=\nu$ can be obtained by duality in a coherent way with the double duality test, {\it on the condition to know what sequence must be 
 used}. \\
However, we have already proved in ([36],[38],[41],[42],[44]) that the {\it Cauchy stress equations} must be replaced by the {\it Cosserat couple-stress equations} and that the {\it Janet sequence} (only used in this paper) must be thus replaced by the {\it Spencer sequence}. Accordingly, it becomes clear that the work of Lanczos and followers has been based on a {\it double confusion} between fields and inductions on one side, but also between the Janet sequence and the Spencer sequence on the other side. \\
 
\noindent
{\bf FUNDAMENTAL RESULT  6.10}: The Janet and Spencer sequences for {\it any} Lie operator of finite type are formally exact by construction, both with their corresponding adjoint sequences. Lanczos has been trying to parametrize $ad({\cal{D}}_1)$ by $ad({\cal{D}}_2)$ when ${\cal{D}}_1$ parametrizes ${\cal{D}}_2$. On the contrary, we have proved that one must parametrize $ad({\cal{D}})$ by $ad({\cal{D}}_1)$ when ${\cal{D}}$ parametrizes ${\cal{D}}_1$ as in the famous {\it infinitesimal equivalence problem} ([29], p 332-336), {\it with a shift by one step}. \\

 \noindent
 {\bf CONCLUSION}  \\
 
 The effective usefulness of the double duality test seems absolutely magical in actual practice but the reader may not forget about the amount of mathematics needed from different domains. Unhappily, in our opinion based on a long experience in dealing with applications, the most difficult part is concerned with formal integrability and involution needed in order to compute the various differential ranks involved. However, the above methods, which are superseding the pioneering approaches of Janet and Cartan, are still not known in mathematical physics and mechanics or even in control theory despite many tentatives done twenty years ago. We hope that this paper will help improving this situation in a near future, in particular when dealing with {\it partial differential optimal control}, that is with variational calculus with OD or PD constraints along the way that has been initiated by Lanczos for eliminating the corresponding Lagrange multipliers or using them as potentials.  \\
 
\noindent
{\bf REFERENCES}  \\

\noindent
[1] G.B. AIRY: On the Strains in the Interior of Beams, Phil. Trans. Roy. Soc.London, 153, 1863, 49-80. \\
\noindent
[2] F. BAMPI, G. CAVIGLIA: Third-order Tensor Potentials for the Riemann and Weyl Tensors, General Relativity and Gravitation, 15, 1983, 375-386.  \\ 
 \noindent
[3] E. BELTRAMI: Osservazioni sulla Nota Precedente, Atti Reale Accad. Naz. Lincei Rend., 5, 1892, 141-142.  \\
 \noindent
[4] J.E. BJORK: Analytic D-Modules and Applications, Kluwer, 1993.\\
\noindent
[5] N. BOURBAKI: El\'{e}ments de Math\'{e}matiques, Alg\`{e}bre, Ch. 10, Alg\`{e}bre Homologique, Masson, Paris, 1980.   \\
\noindent
[6] E. CARTAN: Les Syst\`{e}mes Diff\'{e}rentiels Ext\'{e}rieurs et Leurs Applications G\'{e}om\'{e}triques, Hermann, Paris,1945.  \\
\noindent 
[7] S.B. EDGAR: On Effective Constraints for the Riemann-Lanczos Systems of Equations. J. Math. Phys., 44, 2003, 5375-5385. \\ 
 http://arxiv.org/abs/gr-qc/0302014   \\
\noindent
[8] S.B. EDGAR, A. H\"{O}GLUND: The Lanczos Potential for the Weyl Curvature Tensor: Existence, Wave Equations and Algorithms, Proc. R. Soc. Lond., A, 453, 1997, 835-851.  \\
\noindent
[9] S.B. EDGAR, A. H\"{O}GLUND: The Lanczos potential for Weyl-Candidate Tensors Exists only in Four Dimension, General Relativity and Gravitation, 
 32, 12, 2000, 2307. \\
http://rspa.royalsocietypublishing.org/content/royprsa/453/1959/835.full.pdf   \\ 
 \noindent
[10] S.B. EDGAR, J.M.M. SENOVILLA: A Local Potential for the Weyl tensor in all dimensions, Classical and Quantum Gravity, 21, 2004, L133.\\
 http://arxiv.org/abs/gr-qc/0408071     \\
\noindent
[11] A. EINSTEIN: Die Feldgleichungen der Gravitation, Sitz. Preus. Akad. der Wiissenschaften zu Berlin, 1915, 844-847.  \\
\noindent
[12] L.P. EISENHART: Riemannian Geometry, Princeton University Press, 1926.  \\
\noindent
[13] M. JANET: Sur les Syst\`{e}mes aux D\'{e}riv\'{e}es Partielles, Journal de Math., 8(3),1920, 65-151.  \\
\noindent
[14] E.R. KALMAN, Y.C. YO, K.S. NARENDA: Controllability of Linear Dynamical Systems, Contrib. Diff. Equations, 1, 2, 1963, 189-213.\\
\noindent
[15] M. KASHIWARA: Algebraic Study of Systems of Partial Differential Equations, M\'emoires de la Soci\'et\'e Math\'ematique de France 63, 1995, 
(Transl. from Japanese of his 1970 Master's Thesis).\\
\noindent
[16] A. KUMPERA, D.C. SPENCER, Lie Equations, Ann. Math. Studies 73, Princeton University Press, Princeton, 1972.\\
\noindent
[17] E. KUNZ: Introduction to Commutative Algebra and Algebraic Geometry, Birkh\"{a}user, Boston, 1985.  \\
\noindent
[18] C. LANCZOS: Lagrange Multiplier and Riemannian Spaces, Reviews of Modern Physics, 21, 1949, 497-502.  \\ 
\noindent
[19] C. LANCZOS: The Splitting of the Riemann Tensor, Rev. Mod. Phys. 34, 1962, 379-389.  \\
\noindent
[20] C. LANCZOS: The Variation Principles of Mechanics, Dover, New York, 4th edition, 1949.  \\
\noindent
[21] F. S. MACAULAY, The Algebraic Theory of Modular Systems, Cambridge Tracts 19, Cambridge University Press, London, 1916; 
Reprinted by Stechert-Hafner Service Agency, New York, 1964.\\
\noindent
[22] E. MASSA, E. PAGANI: Is the Rieman Tensor Derivable from a Tensor Potential, Gen.Rel. Grav., 16,1984, 805-816.  \\ 
\noindent
[23] J.C. MAXWELL: On Reciprocal Figures, Frames and Diagrams of Forces, Trans. Roy. Soc. Ediinburgh, 26, 1870, 1-40. \\
\noindent
[24] G. MORERA: Soluzione Generale della Equazioni Indefinite dell'Equilibrio di un Corpo Continuo, Atti. Reale. Accad. dei Lincei, 1, 1892, 137-141+233.  \\
\noindent
[25] U. OBERST: Multidimensional Constant Linear Systems, Acta Appl. Math., 20, 1990, 1-175.   \\
\noindent
[26] U. OBERST: The Computation of Purity Filtrations over Commutative Noetherian Rings of Operators and their Applications to Behaviours, Multidim. Syst. Sign. Process. (MSSP), Springer, 2013.\\
http://dx.doi.org/10.1007/s11045-013-0253-4  \\
\noindent
[27] P. O'DONNELL, H. PYE: A Brief Historical Review of the Important Developments in Lanczos Potential Theory, EJTP, 24, 2010, 327-350.  \\
\noindent
[28] V.P. PALAMODOV: Linear Differential Operators with Constant Coefficients, Grundlehren der Mathematischen Wissenschaften 168, Springer, 1970.\\
\noindent
[29] J.-F. POMMARET: Systems of Partial Differential Equations and Lie Pseudogroups, Gordon and Breach, New York, 1978 
(Russian translation by MIR, Moscow, 1983) \\
\noindent
[30]  J.-F. POMMARET: Lie Pseudogroups and Mechanics, Gordon and Breach, New York, 1988.\\
\noindent
[31] J.-F. POMMARET: Partial Differential Equations and Group Theory,New Perspectives for Applications, Mathematics and its Applications 293, Kluwer, 1994.\\
http://dx.doi.org/10.1007/978-94-017-2539-2   \\
\noindent
[32] J.-F. POMMARET: Partial Differential Control Theory, Kluwer, 2001, 957 pp.\\
\noindent
[33] J.-F. POMMARET: Algebraic Analysis of Control Systems Defined by Partial Differential Equations, in Advanced Topics in Control Systems Theory, Lecture Notes in Control and Information Sciences 311, Chapter 5, Springer, 2005, 155-223.\\
\noindent
[34] J.-F. POMMARET: Gr\"{o}bner Bases in Algebraic Analysis: New perspectives for applications, Radon Series Comp. Appl. Math 2, 1-21, de Gruyter, 2007.\\
\noindent
[35] J.-F. POMMARET: Arnold's hydrodynamics revisited, AJSE-mathematics, 34, 1, 2009, 157-174).  \\
https://ajse-mathematics.kfupm.edu.sa/Old/May2009.asp    \\
\noindent
[36] J.-F. POMMARET: Parametrization of Cosserat Equations, Acta Mechanica, 215, 2010, 43-55.\\
\noindent
[37] J.-F. POMMARET: Macaulay Inverse Systems Revisited, Journal of Symbolic Computation, 46, 2011, 1049-1069.  \\
http://dx.doi.org/10.1016/j.jsc.2011.05.007    \\
\noindent
[38] J.-F. POMMARET: Spencer Operator and Applications: From Continuum Mechanics to Mathematical Physics, in "Continuum Mechanics-Progress in Fundamentals and Engineering Applications", Dr. Yong Gan (Ed.), ISBN: 978-953-51-0447--6, InTech, 2012, Chapter 1, Available from: \\
http://www.intechopen.com/books/continuum-mechanics-progress-in-fundamentals-and-engineering-applications   \\
\noindent
[39] J.-F. POMMARET: Deformation Cohomology of Algebraic and Geometric Structures, arXiv:1207.1964.  \\
http://arxiv.org/abs/1207.1964  \\
\noindent
[40] J.-F. POMMARET: Relative Parametrization of Linear Multidimensional Systems, Multidim. Syst. Sign. Process. (MSSP), Springer, 26, 2013, 405-437. \\
http://dx.doi.org/10.1007/s11045-013-0265-0  \\
\noindent
[41] J.-F. POMMARET: The Mathematical Foundations of General Relativity Revisited, Journal of Modern Physics, 4, 2013, 223-239.\\
http://dx.doi.org/10.4236/jmp.2013.48A022   \\
\noindent
[42] J.-F. POMMARET: The Mathematical Foundations of Gauge Theory Revisited, Journal of Modern Physics, 2014, 5, 157-170.  \\
http://dx.doi.org/10.4236/jmp.2014.55026    \\
\noindent
[43] J.-F. POMMARET: Macaulay Inverse Systems and Cartan-Kahler Theorem, arXiv:1411.7070  \\
http://arxiv.org/abs/1411.7070  \\
\noindent
[44] J.-F. POMMARET: From Thermodynamics to Gauge Theory: the Virial Theorem Revisited, in " Gauge Theories and Differential geometry ", NOVA Science Publishers, 
2015, Chapter 1, 1-44.  \\
http://arxiv.org/abs/1504.04118  \\
\noindent
[45] J.-F. POMMARET, A. QUADRAT: Localization and parametrization of linear multidimensional control systems, Systems \& Control Letters 37, 1999, 247-260.  \\
\noindent
[46] J.-F. POMMARET, A. QUADRAT: Algebraic Analysis of Linear Multidimensional  Control Systems, IMA Journal of Mathematical Control and Informations, 
16, 1999, 275-297.\\
\noindent
[47] A. QUADRAT, D. ROBERTZ: A Constructive Study of the Module Structure of Rings of Partial Differential Operators, Acta Applicandae Mathematicae, 
133, 2014, 187-234. \\
http://hal-supelec.archives-ouvertes.fr/hal-00925533  \\
\noindent
[48] C. RIQUIER: Les Syst\`{e}mes d'Equations aux D\'{e}riv\'{e}es Partielles, Gauthiers-Villars, Paris, 1910.  \\
\noindent
[49] M.D. ROBERTS: The Physical Interpretation of the Lanczos Tensor, Il Nuovo Cimento, B110, 1996, 1165-1176.  \\ 
http://arxiv.org/abs/gr-qc/9904006   \\   
\noindent
[50] J.J. ROTMAN: An Introduction to Homological Algebra, Pure and Applied Mathematics, Academic Press, 1979.\\
\noindent
[51] J.-P. SCHNEIDERS: An Introduction to D-Modules, Bull. Soc. Roy. Sci. Li\`{e}ge, 63, 1994, 223-295.  \\
\noindent
[52] D.C. SPENCER: Overdetermined Systems of Partial Differential Equations, Bull. Amer. Math. Soc., 75, 1965, 1-114.\\
\noindent
[53] P.P. TEODORESCU: Stress Functions in 3-Dimensional Elastodynamics, Acta Meh., 14, 1972, 103.  \\
\noindent
[54] E. ZERZ: Topics in Multidimensional Linear Systems Theory, Lecture Notes in Control and Information Sciences 256, Springer, 2000.\\

\end{document}